\documentclass[12pt]{iopart}

\usepackage{iopams}  
\usepackage[T1]{fontenc}
\usepackage{url} 
\usepackage{makeidx} 
\usepackage[pdftex]{graphicx} 
\usepackage{amsfonts} 
\expandafter\let\csname equation*\endcsname\relax
 \expandafter\let\csname endequation*\endcsname\relax
 \usepackage{amsmath} 
\usepackage{listings} 
\usepackage{verbatim}
\usepackage{cite}
\usepackage{hyperref} 
\usepackage{array} 
\usepackage[bold]{hhtensor}
\usepackage{geometry}
\hypersetup{ 
    pdftitle={Technical Notes},    
    colorlinks=false,       
    linkcolor=red,          
    citecolor=green,        
    filecolor=magenta,      
    urlcolor=cyan           
}
\usepackage{graphicx}
\usepackage{caption}
\usepackage{subcaption}
\usepackage{movie15}
\usepackage{color}
\usepackage[table]{xcolor}
\usepackage{booktabs}

\begin{document}

\title[Sandoval, Campbell, Marian]{Thermodynamic interpretation of reactive processes in Ni-Al nanolayers from atomistic simulations}

\author{Luis Sandoval \footnote{Currently at Theoretical Division T-1,
    Los Alamos National Laboratory, Los Alamos, NM 87545, USA.}, 
Geoffrey H.\ Campbell, Jaime Marian}
\address{Lawrence Livermore National Laboratory}

\begin{abstract}
Metals which can form intermetallic compounds by exothermic reactions constitute a class of reactive materials with multiple applications. Ni-Al laminates of thin alternating layers are being considered as model nanometric metallic multilayers for studying various reaction processes. However, the reaction kinetics at short timescales after mixing are not entirely understood. In this work, we calculate the free energies of Ni-Al alloys as a function of composition and temperature for different solid phases using thermodynamic integration based on state-of-the-art interatomic potentials. We use this information to interpret molecular dynamics (MD) simulations of bilayer systems at 800 K and zero pressure, both in isothermal and isenthalpic conditions. 
We find that a disordered phase always forms upon mixing as a precursor to a more stable nano crystalline B2 phase. We construe the reactions observed in terms of thermodynamic trajectories governed by the state variables computed. Simulated times of up to 30 ns were achieved, which provides a window to phenomena not previously observed in MD simulations. Our results provide insight into the early experimental reaction timescales and suggest that the path (segregated reactants)$\rightarrow$(disordered phase)$\rightarrow$(B2 structure) is always realized irrespective of the imposed boundary conditions. 

\end{abstract}

\maketitle

\section{Introduction}\label{sec:intro}
Solid reactive materials that form intermetallic compounds via high energy release are increasingly being used in multiple materials science processes \cite{nuzzo1986,intro1,roga2008}. These systems have the advantage that both reactants and products are confined to the condensed state, which makes them helpful in anaerobic conditions and where gaseous products are non-desirable. For these reasons, there is a wide range of applications where they are now being used, such as welding, propellants, heat initiators, etc. Ni-Al systems are an important subclass of reactive materials due to the formation of intermetallic phases with high temperature strength and high resistance to oxidation \cite{barmak1997,gunduz2009}. 

Because of their many attractive properties and promising applications, Ni-Al reactive systems have attracted significant attention over the last two decades, both experimental \cite{kim2008,kim2011,trenkle2010,swa2013} and theoretical \cite{salloum2010,vohra2011}. The formation of Ni-Al intermetallics is an intrinsically atomistic process, governed by the free energies of the different phases involved as well as the kinetics of atomic and interfacial motion. With the advent of efficient simulation codes and reliable interatomic potentials, molecular dynamics (MD) has emerged as an ideal tool to investigate these reactive processes. However, despite the number of MD studies performed to date \cite{geyser2000,yu2007,delogu,ev2011}, a comprehensive thermodynamic picture of the reaction processes is still lacking. In this paper, we use state-of-the-art interatomic potentials to calculate the free energies of different Ni-Al phases. We then use this thermodynamic information to interpret and understand the reactive behavior of Ni-Al nano laminates. We carry out simulations at 800 K, which is known to be above the temperature of self ignition in NiAl ($\approx$600 K), and zero pressure, both under isothermal and isenthalpic conditions. We find that, in both cases, a recrystallized B2 phase forms from a local disordered phase (which can be considered the liquid phase or a local amorphous phase, depending on the situation) caused primarily by the swift penetration of Ni into the Al layer.

This paper is organized as follows. In Section \ref{sec:methods}, we describe the simulation method in detail, as well as the thermodynamic integration technique employed. We then begin Section \ref{sec:results} providing the energy vs.~volume curves for several solid phases of equiatomic Ni-Al, followed by the internal and free energy results as a function of alloy composition, and a description of Ni-Al bicrystal reaction kinetics. We finalize in Section \ref{sec:disc} with a discussion of our findings and the conclusions. 

\section{Methods}\label{sec:methods}
All calculations presented here were carried out with the \verb+lammps+ code \cite{lammps} using 256 processors on Livermore Computing's parallel architectures. 
We employ the embedded-atom method (EAM) interatomic potential for Ni-Al developed by Purja and Mishin \cite{mishin2009}. This potential is particularly suitable for simulations of heterophase interfaces and mechanical behavior of Ni-Al alloys. Additionally, the melting points of fcc Ni ---1700 K--- and Al ---1040 K--- are well reproduced by the potential, which adds confidence to the high temperature calculations that will be presented here. Except where noted, we consider periodic systems at zero total pressure.

To compute free energies we use thermodynamic integration using Kirkwood's coupling parameter method n                                                                                                                                                                                                                                                                                                                                                                                                                                                                                                                                                                                                                     \cite{kirkwood1935,rickman2002,muller2007,tuckerman2008}, also known as $\lambda$-\emph{integration}. In the canonical ensemble, the free energy difference, $\Delta F=F_{B}-F_{A}$, between two systems $A$ and $B$ characterized by potential energy functions $U_A$ and $U_B$ can be obtained by integrating along a reversible path from $A$ to $B$. The distance along this path can be measured by using a potential energy function that uses a switching parameter $\lambda$:
\begin{equation}
U(\lambda)=(1-\lambda)U_{A}+\lambda U_{B}
\end{equation}
The canonical partition function for such a system can be written as: 
\begin{equation}
Q(N;\Omega;T;\lambda)=C\int{d\vec{r}^N}\exp{\left\{-\beta U(\lambda)\right\}}
\label{eq:q}
\end{equation}
where $N$ is the number of particles, $\Omega$ is the system volume, $T$ is the absolute temperature, and $C$ is a constant that includes the result of integration over momenta and other physical constants. $\beta=(k_BT)^{-1}$ is the reciprocal temperature with $k_B$ being Boltzmann's constant. From eq.\ \ref{eq:q}  the Helmholtz free energy can be calculated as $F=-\beta^{-1}\ln{Q}$, whose derivative with respect to the switching parameter can be written as:
\begin{equation}
\left.\frac{\partial F(\lambda)}{\partial\lambda}\right|_{N,\Omega,T}=-\frac{1}{\beta}\frac{\partial}{\partial\lambda}\ln{Q}=
-\frac{1}{\beta Q}\frac{\partial Q}{\partial\lambda}=\frac{\int{d\vec{r}^N}\frac{\partial U(\lambda)}{\partial\lambda}\exp{\left\{-\beta U(\lambda)\right\}}}{\int{d\vec{r}^N}\exp{\left\{-\beta U(\lambda)\right\}}}
\label{eq:deriv}
\end{equation}
which is the expression of an ensemble average that can be calculated via molecular dynamics simulations. From this, the free energy difference between systems $A$ and $B$ is given by:
\begin{equation}
\Delta F=F(U_A)-F(U_B)=\int_0^1\left\langle\frac{\partial U}{\partial\lambda}\right\rangle d\lambda=\int_0^1\left\langle U_{B}-U_A\right\rangle d\lambda
\label{int}
\end{equation}	 

For solid phases with ordered crystal structure, a system of Einstein oscillators is appropriate as a reference state (characterized by $U_B$). The Einstein crystal can also be used for systems without short-range order such as amorphous phases provided that internal transport processes (e.g. self-diffusion) are negligible on the scale of the simulations:
\begin{equation}
U_B(\vec{r})=\frac{\alpha}{2}\sum_{i=1}^N{\left(\vec{r}_i-\vec{r}_{0,i}\right)^2}
\label{einstein}
\end{equation}
where $\alpha$ is a spring constant\footnote{For numerical reasons, it is best to choose $\alpha$ in eq.\ \ref{einstein} such that:
$$\alpha=\frac{3}{\beta\langle\Delta r^2\rangle}$$
where $\Delta r^2$ is the mean square displacement of the target phase.} and $\{\vec{r}_0\}$ are the equilibrium positions. After accounting for the use of periodic boundary conditions and fixing the center of mass in MD simulations, the free energy of a system of harmonic oscillators can be obtained analytically as:
\begin{equation}
F(U_B)=-\frac{3N}{2\beta}\left\{
\log{\frac{m\alpha^{N-1}\beta^{N-2}}{h^2}}+\frac{1}{3N}\log{\frac{N}{\Omega^2}}
\right\}
\label{fuo2}
\end{equation}
where $m$ and $h$ are, respectively, the atomic mass and Planck's constant.

The free energy of the system at temperatures other than the reference temperature, $T_0$, used in the above process can be found by recourse to the Gibbs-Helmholtz integral: 
\begin{equation}
\frac{F}{T}=\frac{F_0}{T_0}-\int_{T_0}^{T}{\frac{E(n,\Omega,\theta)}{\theta^2}d\theta}
\label{hg}
\end{equation}
where $F_0$ is the free energy at $T_0$ and $E$ is the internal energy. $E$ can be computed as the ensemble average of the total energy, decomposed into potential and kinetic energies: 
\begin{equation}
E(T)=\langle U(T)+K(T)\rangle
\label{et}
\end{equation}

\begin{figure}[h]
        \centering
                \includegraphics[width=1.5\linewidth,trim=1cm 2.0cm 0cm 16.0cm,clip=true]{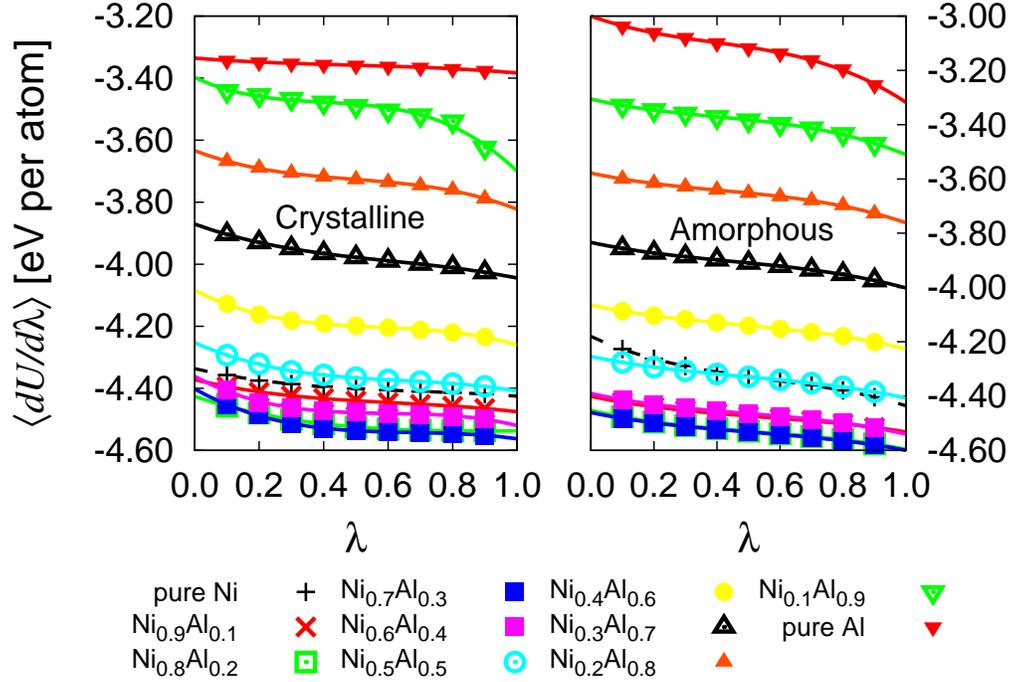}
         \caption{Calculation of the integrand in eq.\ \ref{int} as a function of $\lambda$ for several alloy compositions in random bcc (crystalline) and amorphous Ni-Al phases at 500 K. The curves represent cubic polynomial fits to the data points. Integration of these curves as in eq.\ \ref{int} yields the free energy of the system. In the crystalline case, the pure Ni and Al phases correspond to fcc crystals, while a bcc phase is considered for mixed compounds with given atomic compositions. Note that the vertical scale is different in both graphs.}\label{fig:lambda}
\end{figure}

By way of example, Figure \ref{fig:lambda} shows the calculation of the integrand in eq.\ \ref{int} as a function of $\lambda$ for several alloy compositions in random bcc  and amorphous Ni-Al phases at 500 K. The curves represent cubic polynomial fits to the data points. Integration of these curves as in eq.\ \ref{int} yields the free energy of the system. Sufficient sampling is critical, particularly when $\langle\partial U/\partial\lambda\rangle$ changes quickly, to ensure an accurate calculation of the free energy integral.
To predict critical (phase transition) points with accuracy, $\lambda$ must be resolved very finely near $\lambda=1$ (in 0.01 intervals), which makes these calculations very costly. Methods to mitigate this difficulty have been discussed in the literature \cite{resat1993}.

\begin{figure}[h]
        \centering
                \includegraphics[width=1.5\linewidth,trim=1cm 2.0cm 0cm 16.0cm,clip=true]{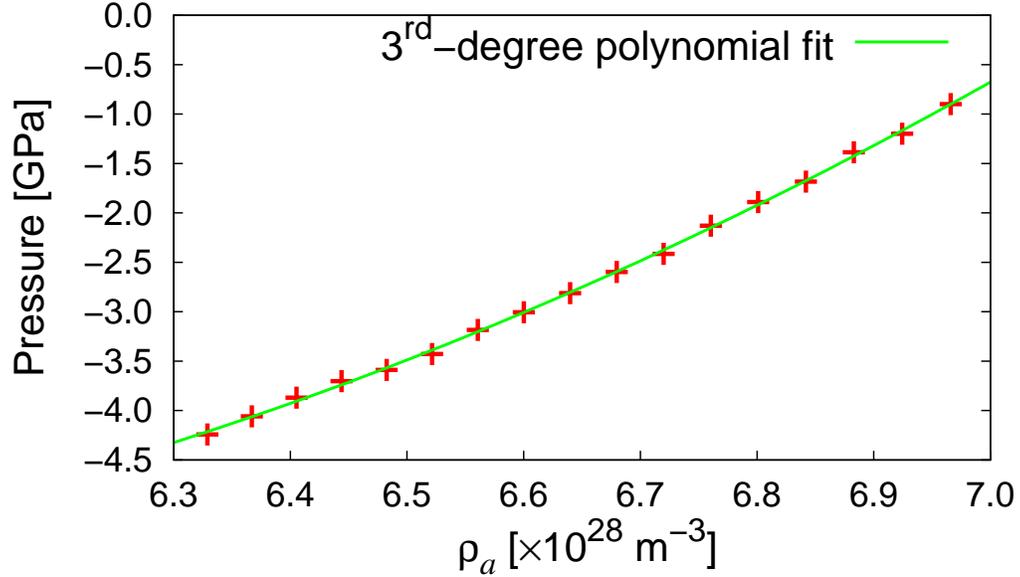}
         \caption{Equation of state for a liquid NiAl mixture at 2000 K.}\label{fig:ideal}
\end{figure}

For liquid systems, defined as those in which diffusion takes place on the order of the simulation time scale, the Einstein crystal can no longer be used as a reference state. In such cases, 
the free energy is obtained as \cite{mei1992,frenkelbook,glosli1999}:
\begin{equation}
F_l=F_{gm}(\rho_0,T_0)+\int_0^{\rho_0}d\rho\left[\frac{p(T_0,\rho)-\rho k_BT_0}{\rho^2}\right]
\label{liquid}
\end{equation}
where $p$ is the pressure, $\rho_0$ is the system's density at a temperature $T_0$ above the supercritical temperature (here we have taken $T_0=2000$ K) and $F_{gm}(\rho,T)$ is the free energy of a binary ideal gas mixture at density and temperature $\rho$ and $T$:
$$F_{gm}(\rho,T)=\frac{1}{2}\left[F_g^{Ni}(\rho,T)+F_g^{Al}(\rho,T)\right]+k_BT\ln\frac{1}{4}$$
with \cite{broughton1997}:
$$F_g^i(\rho,T)=N_ik_BT\left\{\ln\left[\rho\left(\frac{h^2}{2\pi m_ik_BT}\right)^{3/2}\right]-1\right\}$$
The integrand in the second term of the r.h.s.~of eq.\ \ref{liquid} represents an isothermal expansion from $\rho_0$ to zero density (i.e.~infinite volume), where the system effectively behaves as an ideal gas. This expansion should be reversible, which means that no first-order transition ---e.g.~liquid-gas--- should be traversed. The equation of state $p(T_0,\rho)$ is obtained from a set of canonical ensemble calculations at $T_0$ of an equiatomic liquid mixture of Ni and Al. This is shown in Figure \ref{fig:ideal}. The resulting data are fitted to a 3$^{\rm{rd}}$ degree polynomial and the integral in eq.\ \ref{liquid} is solved to yield the free energies.

\section{Results}\label{sec:results}

\subsection{Equiatomic NiAl}
First we present results for equiatomic systems (Ni$_1$Al$_1$) containing 16,000 atoms. We study four distinct phases, namely, the ordered fcc L1$_0$, ordered (B2) and disordered (random solid solution) bcc phases, and an amorphous phase obtained from quenching a liquid system (equilibrated for 100 ps at 3000 K) at a cooling rate of 300 K ps$^{-1}$ \cite{noya2002}. We note that the term `amorphous' is employed loosely in this paper to refer to an unstructured material, be it in the solid or in the liquid state. Typically an amorphous system can always be considered the liquid structure of a certain solid phase, although it may not necessarily correspond to the absolutely lowest free energy structure.

\subsubsection{Equilibrium volume and thermal expansion.}
\begin{figure}[h]
        \centering
                \includegraphics[width=\textwidth,trim=1cm 1.0cm 6cm 15.0cm,clip=true]{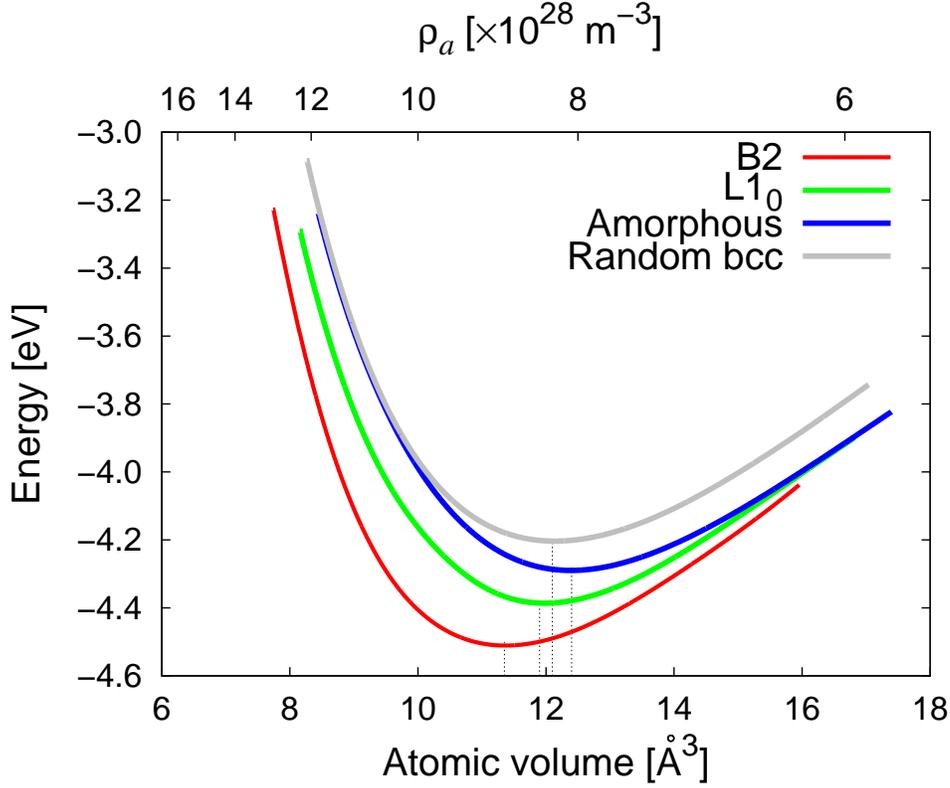}
                \caption{Cohesive energy of four different equiatomic Ni-Al phases as a function of atomic volume $\Omega_a$ (or, equivalently, atomic density $\rho_a$). The vertical dashed lines indicate the location of the minimum for each curve, given in Table \ref{table}.}
                \label{eos}
\end{figure}

The thermal expansion coefficient $\alpha_t(T)$ is obtained from the temperature dependence of the atomic volume $\Omega_a$:
\begin{equation}
\alpha_t=\frac{1}{\Omega_0}\frac{d\Omega_a}{dT}
\label{eq:alpha}
\end{equation}
where $\Omega_0$ is a reference atomic volume (usually taken as the value at 0 K). We first compute $\Omega_0$ for the four phases indicated above from energy-volume relations. The evolution of the cohesive energy as a function of atomic volume (and density) is shown in Figure \ref{eos} for each phase. From the figure, one can obtain the equilibrium values as those corresponding to a minimum of the cohesive energy (indicated by vertical dashed lines in Fig.\ \ref{eos}). The numerical values in each case are given in Table \ref{table}. The figure also gives the relative stability of each phase at zero temperature, with the B2 phase always being the most stable.
 
\begin{figure}[h]
         \centering
                \includegraphics[width=\textwidth,trim=1cm 1.0cm 6cm 15.0cm,clip=true]{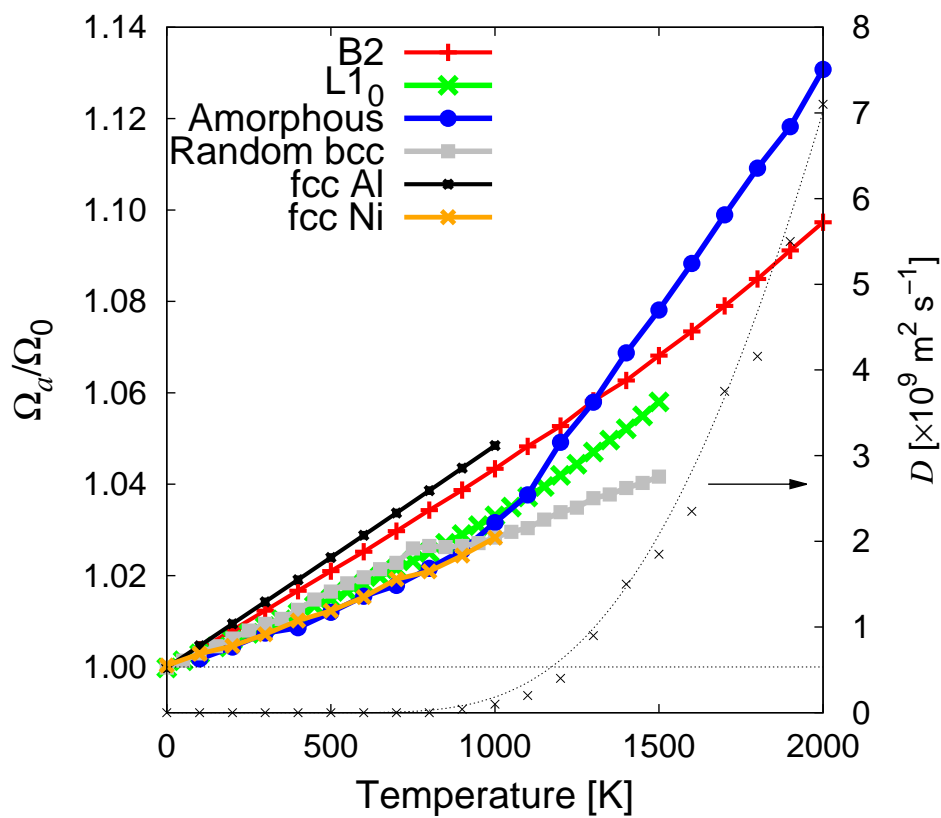}           
\caption{Variation of the atomic volume with temperature for the four different equiatomic Ni-Al phases considered here. The reference atomic volumes $\Omega_0$ are those given in Table \ref{table} for each case. The sharp change in slope for the amorphous system around 1000 K correlates with an increase in self-diffusivity (right-hand axis).}
\label{alpha}
\end{figure}
\begin{table}
\center
\caption{Equilibrium volumes, densities, and lattice parameters at 0 K for the four equiatomic Ni-Al phases considered here (from Fig.\ \ref{eos}). The lattice parameter has been obtained assuming bcc lattice cells for the B2 and solid solution phases and an fcc lattice for the L$1_0$ structure. The amorphous phase does not have an associated underlying lattice structure and hence no equilibrium lattice parameter has been calculated. The last two columns show the values of the thermal expansion coefficient and heat capacity at 800 K, obtained from Figs.\ \ref{alpha} and \ref{ee}.}
\begin{tabular}{*6c}    \toprule
{\small Solid phase} &{\small $\Omega_0$ [\AA$^3$]} & {\small $\rho_0$ [$\times$10$^{28}$ m$^{-3}$]} & {\small $a_0$ [\AA]} & {\small $\alpha_t$ [$\times10^{-5}$ K$^{-1}$]} & {\small $C_p$ [eV atom$^{-1}$ K$^{-1}$}] \\\midrule
\rowcolor{blue!50} {\small B2}		&	11.4		&	8.81			&		2.83	& 4.53 & $2.8\times10^{-4}$\\
\rowcolor{green!50} {\small L1$_0$}		&	11.9			&	8.40			&		3.62 & 4.15 & $2.8\times10^{-4}$\\
\rowcolor{gray!50} {\small Amorphous}	& 12.4		&	8.06			&		-- & 6.05 & $2.6\times10^{-4}$\\
{\small bcc}	& 12.1		&	8.26			&		2.89 & 2.73 & $2.1\times10^{-4}$\\
\rowcolor{orange!50} {\small fcc Ni} & 10.9 & 9.17 & 3.52 & 2.76 & $2.7\times10^{-4}$  \\
\rowcolor{orange!50} {\small fcc Al} & 16.9 &  5.91 & 4.05 & 5.24 & $2.7\times10^{-4}$  \\\bottomrule
 \hline
\end{tabular}
\label{table}
\end{table}

Next we calculate the variation of the atomic volume with temperature using simulations in the isothermal-isobaric ensemble ($NpT$). 
Results for all four phases considered here are shown in Fig.\ \ref{alpha}. The values of $\alpha_t$ at 800 K are given in Table \ref{table}. For the B2 and L1$_0$ systems, the volume increases linearly with temperature, resulting in a constant $\alpha_t$, while for the bcc solid solution it displays some roughness associated with increased internal diffusion and clustering processes. For the amorphous phase, the evolution of the volume with temperature displays two clearly distinguishable linear regimes with a transition at approximately 1000 K. This results in two different values of $\alpha_t$ above and below that transition. This change is due to the transformation from an amorphous solid with high internal viscosity to a liquid with high internal diffusivity. The diffusion coefficient of the amorphous phase has been calculated as a function of temperature and the results plotted against the right-hand axis in Fig.\ \ref{alpha}. Clearly, the diffusivity follows an Arrhenius behavior, $D(\beta)=D_0\exp\left(-\beta E_m\right)$, with $D_0=2.7\times10^{11}$ m$^2$ s$^{-1}$ and $E_m=0.63$ eV.  On the basis of these results, the free energies of the amorphous phase above 1000 K are calculated using eq.~\ref{liquid}.

\subsubsection{Free energies.}
\begin{figure}[h]
       \centering
                \includegraphics[width=\textwidth,trim=1cm 1.0cm 6cm 15.0cm,clip=true]{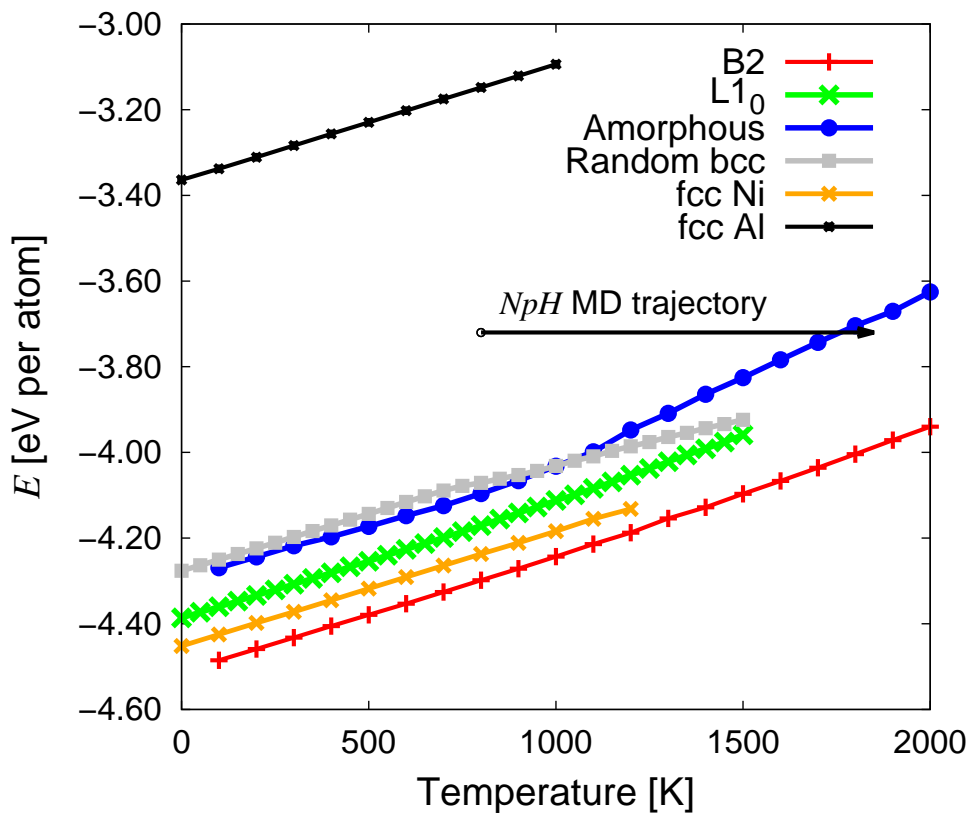}
                \caption{Internal energies as a function of temperature of the four equiatomic Ni-Al phases considered. The internal energies of pure Ni and Al are also shown for reference. The thermodynamic path corresponding to the $NpH$ MD simulations described in Section \ref{nph} is shown.}
       \label{ee}
\end{figure}

\begin{figure}[h]
        \centering
                \includegraphics[width=\textwidth,trim=1cm 1.0cm 6cm 15.0cm,clip=true]{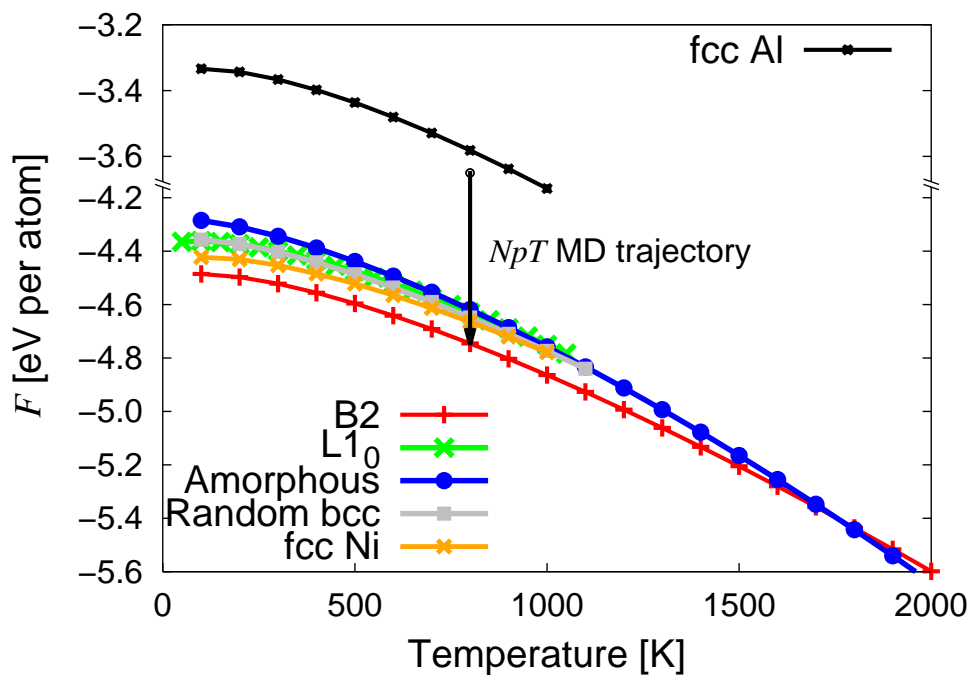}
        \caption{Free energies as a function of temperature of the four equiatomic Ni-Al phases considered. The cut of the free energy curve of the solid phases with that of the amorphous structure yields the melting point in each case. The free energies of pure Ni and Al up to 1000 K are also shown for reference. The thermodynamic path corresponding to the $NpT$ MD simulations described in Section \ref{npt} is shown.}\label{ff}
\end{figure}

From equation \ref{et}, we compute the internal energies as a function of temperature for all the different equiatomic phases as well as for pure fcc Ni and Al for comparison. Results are shown in Figure \ref{ee}, with all systems exhibiting linear dependencies except the amorphous one again at temperatures above 1000 K. From these data, the heat capacity at zero pressure can be calculated straightforwardly as 
$$C_p=\left(\frac{\partial H}{\partial T}\right)_P=\left(\frac{\partial E}{\partial T}\right)_P$$
because at zero pressure $H\equiv E$. The results at 800 K for each phase are given in Table \ref{table}. These calculations are a precursor to obtaining the free energies as a function of temperature (cf.\ eq.\ \ref{hg}), which are given in Figure \ref{ff}. 
Figure \ref{ff} establishes the relative stability of each phase as a function of temperature. The free energies are extended in each case to the crossing temperature with the amorphous phase. This yields the melting points for each of the Ni-Al structures considered: 900 K for the L$1_0$, 1200 K for the random bcc solid solution, and 1780 K for the B2 phase. The latter is the most stable of all the solid phases over the entire temperature range. This has important implications for the reaction kinetics of Ni-Al bilayers that will be studied below.
 
\subsection{Free energies of Ni$_x$Al$_y$ alloys}	
\begin{figure}[h]
        \centering
                \includegraphics[width=1.5\linewidth,trim=1cm 2.0cm 0cm 16.0cm,clip=true]{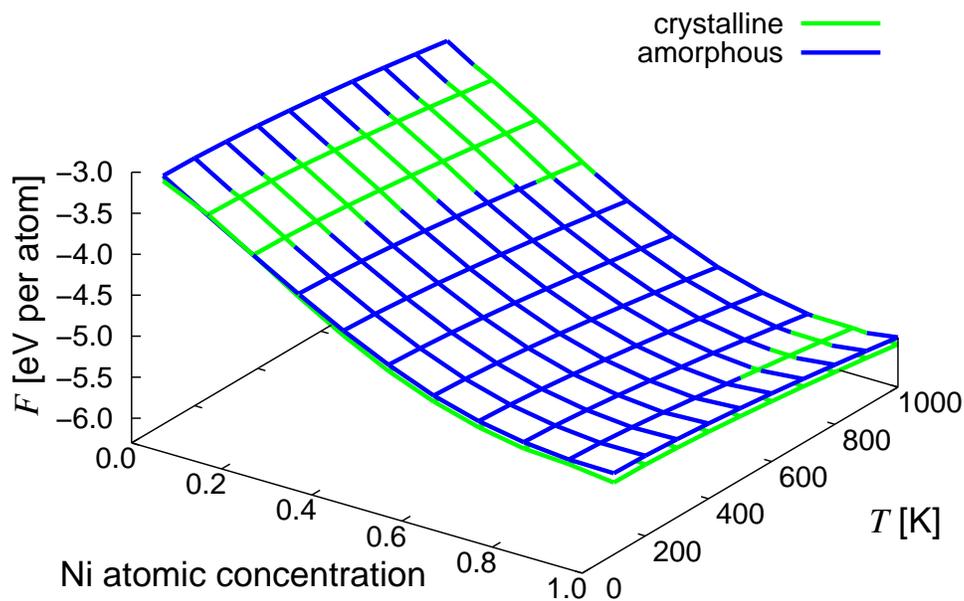}
        \caption{Free energies of amorphous and solid solution bcc phases as a function of temperature and atomic composition. The pure Ni and Al phases ($c$=1.0 and 0.0) have fcc structure.}\label{fig:comp}
\end{figure}	
During the early stages of reactive mixing, the system will probe a wide spectrum of compositions corresponding to different Ni:Al atomic ratios (Ni$_x$Al$_y$). Locally, the $x$:$y$ ratio can be far from unity, depending on fluctuations set by the temperature and the relative diffusivity of each species. This means that the kinetic evolution of the reaction front is set by fluctuations, {\it i.e.}~variations in local composition, temperature, etc. Thus, it is also of interest to calculate the free energy as a function of composition $c$ (which represents the Ni concentration) for selected phases. Since mixing processes involve high entropy and therefore low order, here we focus on the amorphous and solid solution bcc phases as potential precursors of thermodynamically stable (ordered) Ni$_x$Al$_y$ alloys. Here, this amorphous phase can be considered as the liquid structure of the bcc solid solution, although there are lower energy solid structures (e.g.~B2) at these temperatures and, thus, it cannot be considered the absolute liquid configuration. Figure \ref{fig:comp} shows the free energy surface $F(c,T)$ for the two phases of interest. Derivatives of $F(c,T)$ with respect each one of the axes give the entropy (temperature axis) and the chemical potential (concentration axis) differences. The information contained in Fig.\ \ref{fig:comp} can be used to define the phase diagram between these two structures, although neither of these phases are equilibrium phases and thus are typically not considered in Ni-Al phase diagrams.

As the free energy surface shows, the free energy differences between the solid solution bcc and amorphous phases is small, with the crystalline phase generally being more stable than the amorphous one, except roughly for $0.1<c<0.2$, and, interestingly, at $c\approx0.9$ and $T>600$ K. Additionally, the free energies are minimum for Ni concentrations around $c=0.8$. We believe this to be an indirect indicator of the stability of the Ni$_3$Al system.

\subsection{Reaction kinetics}\label{sebsec:kin}
We now study the reactivity of a Ni-Al bilayer at 800 K. The simulated system consists of two crystallites of Ni and Al containing, respectively, $N=313,600$ and 329,251 atoms. The Ni and Al subsystems are fcc lattices oriented in the same direction and separated by an interface with a surface normal oriented along the [100] direction. Periodic boundary conditions are used along each coordinate. Crone \etal~have shown that the ignition temperature (at which a self-sustaining reaction is achieved) depends on the misfit interface strain \cite{crone2011}. To avoid such dependency, we use the columnar arrangement employed by Baras and Politano \cite{baras2011} which ensures a strain-free interface after relaxation. The initial dimensions of the Ni and Al layers are, respectively, 19.6$\times$8.8$\times$19.6 nm, and 17.5$\times$17.5$\times$17.5 nm. This provides for an empty 1-nm thick buffer over which the Al subsystem can expand, relaxing all interfacial stresses.

The bicrystal is initially equilibrated at the target temperature of 800 K by means of an inverse simulated annealing. This is done by heating the system from 0 K at a rate of 1 K ps$^{-1}$ so that it takes 0.8 ns to reach the desired temperature of 800 K. This annealing procedure results in some diffusive mixing prior to reaching 800 K. Therefore, on the Al side of the original interface at $t=0$ , the initial state corresponds to an Al-rich phase with some interpenetrated Ni. This Al-rich phase retains its original fcc order but we have confirmed that it does not correspond to an ordered L1$_0$ structure. On the Ni side of the original interface, the penetration of Al is quite limited, resulting in essentially a very dilute fcc Ni-Al phase. This picture is consistent with relative interdiffusion coefficients that are 3.3 times larger for Ni in Al than vice versa \cite{shankar1978}. 

Experiments are typically conducted at constant pressure and temperature, which suggests running the reaction simulations in the $NpT$ (isobaric-isothermal) ensemble. However, enforcing a constant temperature during typical MD time scales (several nanoseconds) is not representative of a true isothermal process, as the thermostat coupling must be sufficiently strong to extract the released heat over MD time scales. This typically results in unphysically high time relaxation constants that do not reflect the true time scale of the process. 
Thus, the reaction dynamics may be more faithfully simulated using the isobaric-isoenthalpic ensemble $NpH$, where the temperature is free to fluctuate in response to internal transformations under constant enthalpy. For better understanding the reaction kinetics, here we carry out MD simulations from identical initial configurations up to 30 ns under both ensembles. The results for both cases are reported below.

\subsubsection{$NpT$ ensemble}\label{npt}

\begin{figure}[ht]
        \centering
        \begin{subfigure}[h]{0.27\textwidth}
                \includegraphics[width=\textwidth]{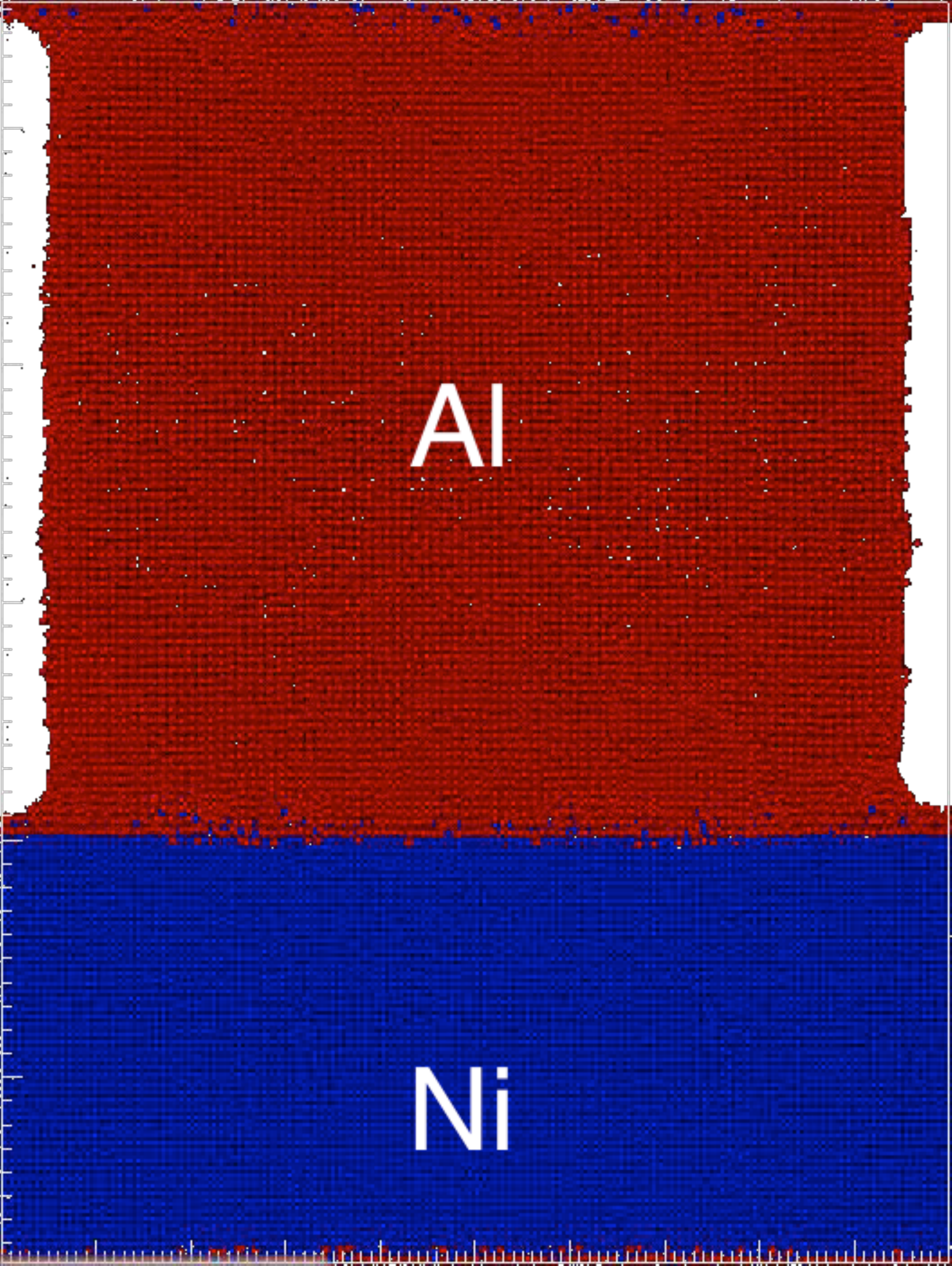}
                \caption{$t=0.0$ ns}
                \label{frame1}
        \end{subfigure}~%
        \begin{subfigure}[h]{0.3\textwidth}
                \includegraphics[width=\textwidth]{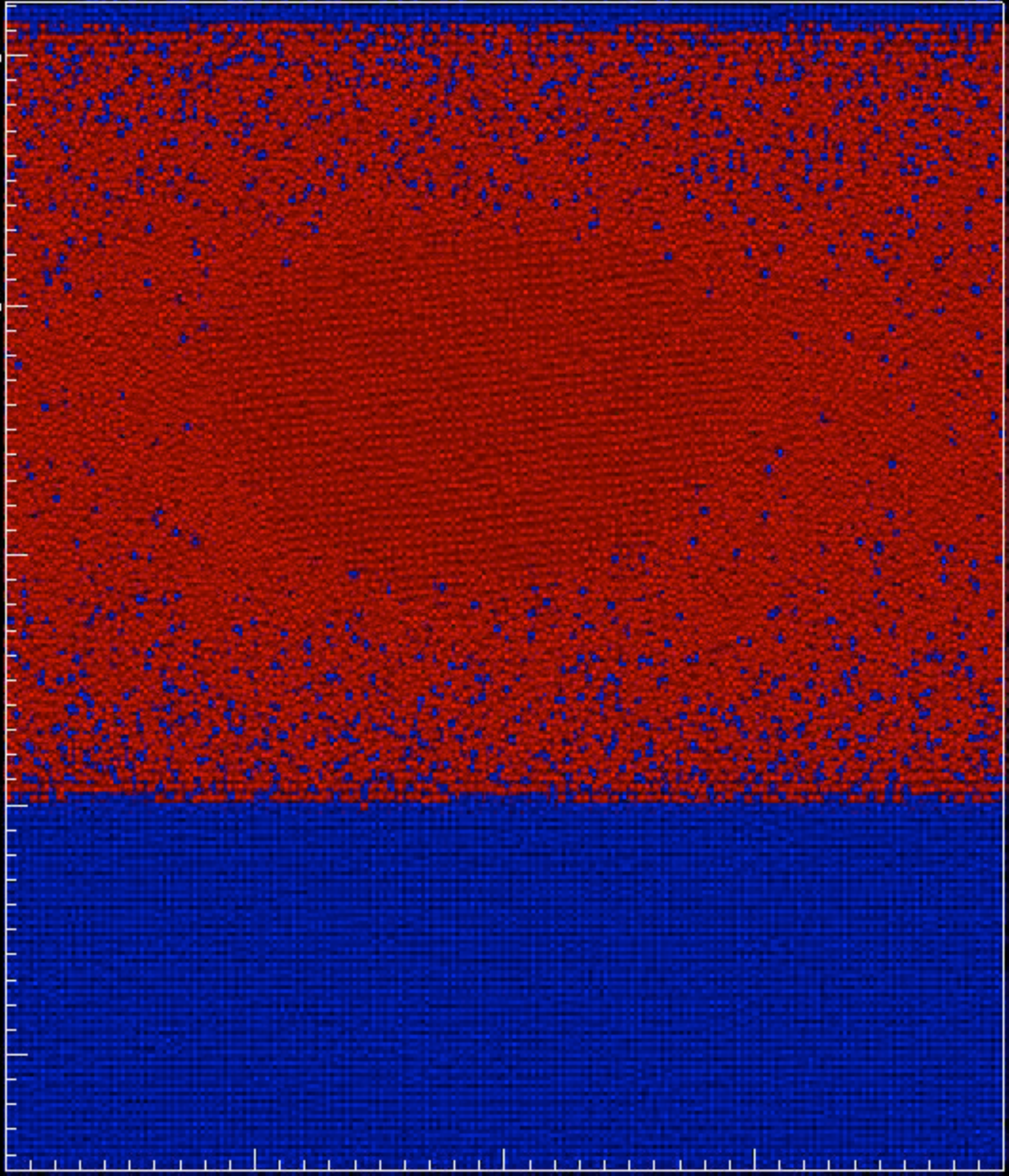}
                \caption{$t=2.5$ ns}
                \label{frame2}
        \end{subfigure}
        \begin{subfigure}[h]{0.3\textwidth}
                \includegraphics[width=\textwidth]{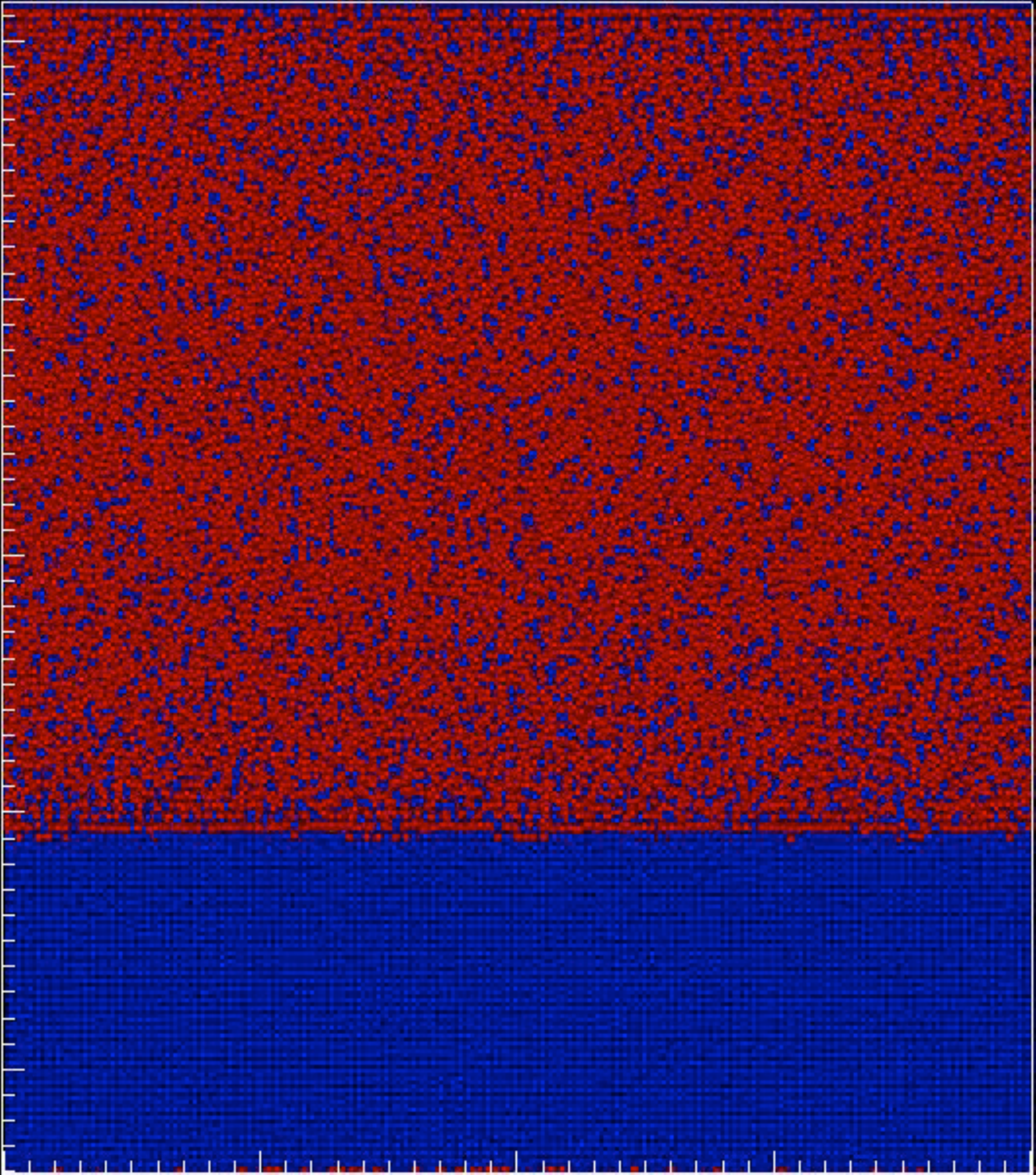}
                \caption{$t=9.0$ ns}
                \label{frame3}
        \end{subfigure}
        \begin{subfigure}[h]{0.3\textwidth}
                \includegraphics[width=\textwidth]{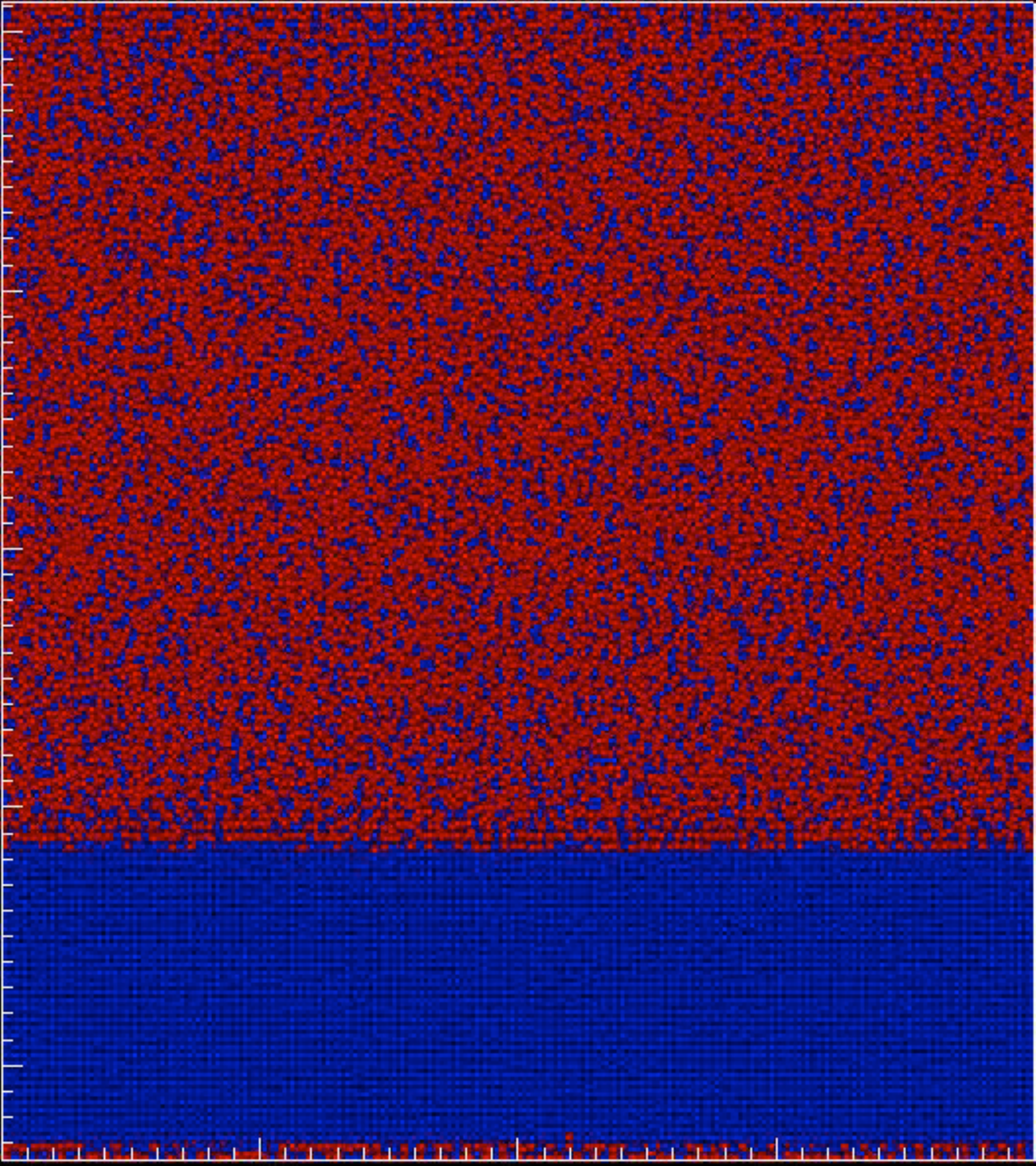}
                \caption{$t=17.5$ ns}
                \label{frame4}
        \end{subfigure}
        \begin{subfigure}[h]{0.3\textwidth}
                \includegraphics[width=\textwidth]{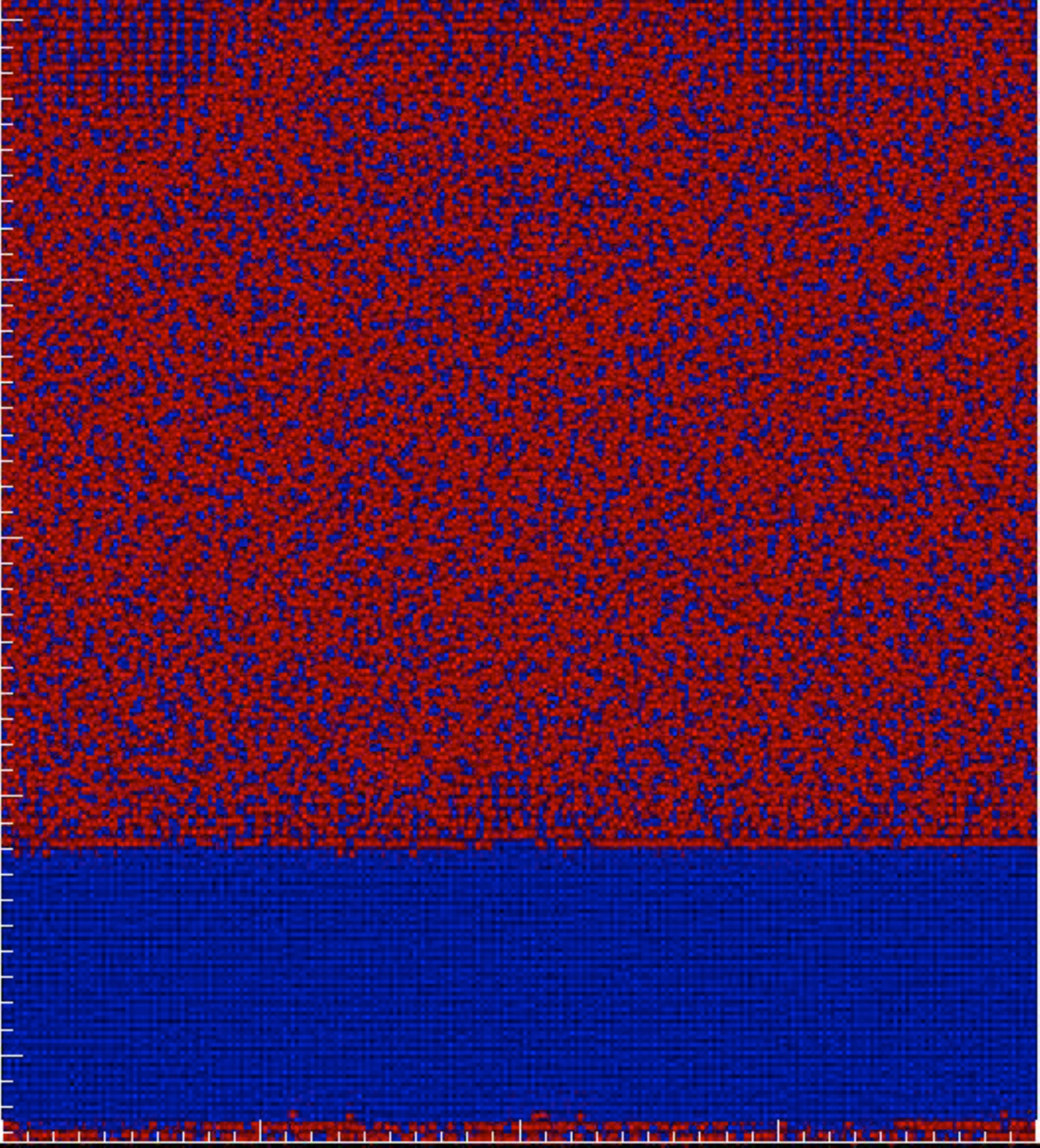}
                \caption{$t=25.5$ ns}
                \label{frame5}
        \end{subfigure}
        \begin{subfigure}[h]{0.3\textwidth}
                \includegraphics[width=\textwidth]{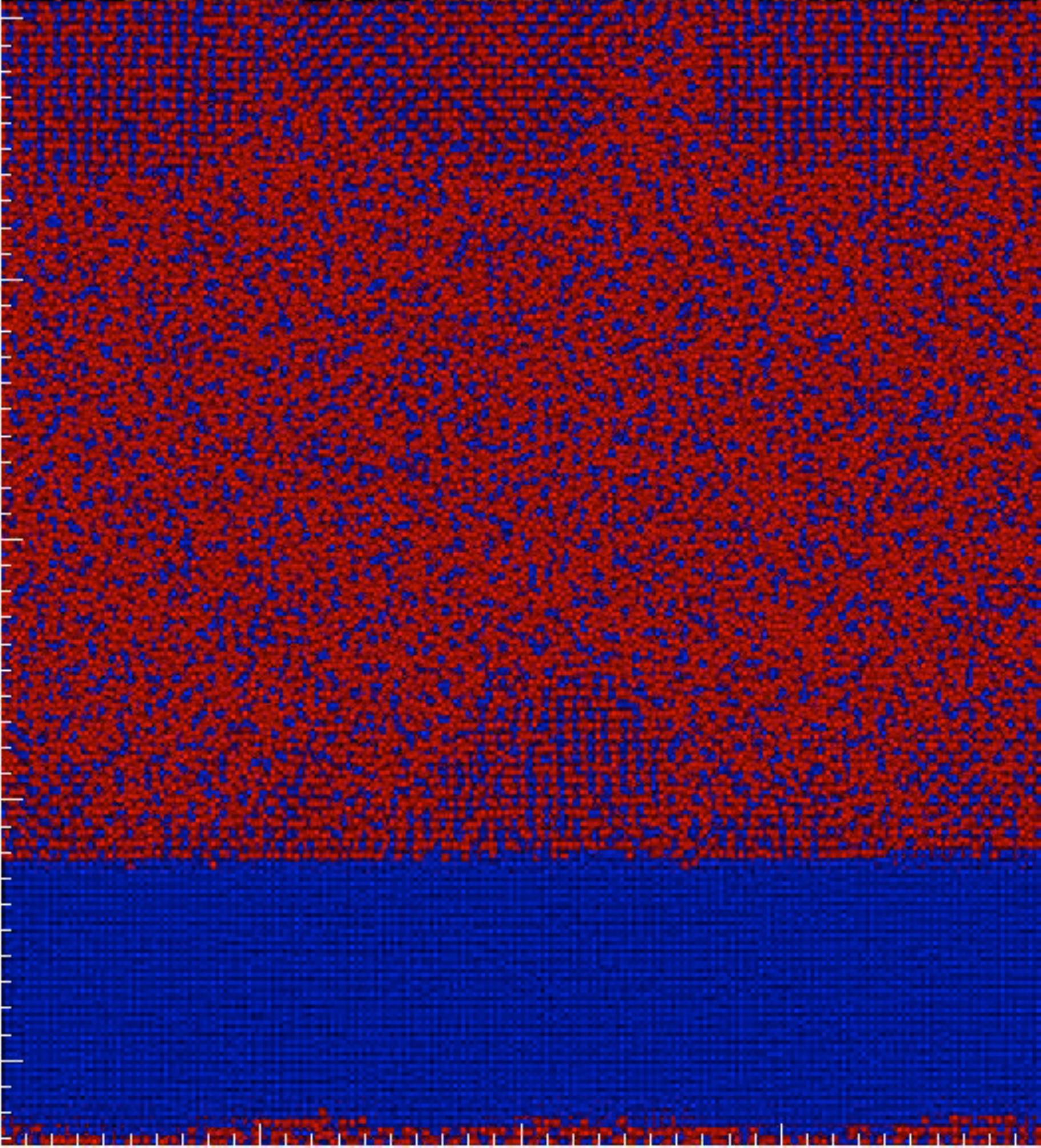}
                \caption{$t=30.0$ ns}
                \label{frame6}
        \end{subfigure}
        \caption{Time sequence of the reaction process in a Ni-Al bilayer at 800 K in the $NpT$ ensemble. Red circles represent Al atoms, blue circles symbolize Ni atoms. (a) Initial system during equilibration. (b) Beginning of the mixing process. Ni penetrates into Al much more than vice versa. (c) Mixing process is nearly complete, amorphization starts. (d) Mixing complete, all amorphous. (e) Crystallization in a B2 (bcc) phase starts at the interface. Crystallization starts where the local composition is close to equiatomic. (f) Crystallization proceeds.}
\label{fig:sequence}
\end{figure}

\begin{figure}[h]
        \centering
                \includegraphics[width=1.5\linewidth,trim=1cm 1.0cm 0cm 15.0cm,clip=true]{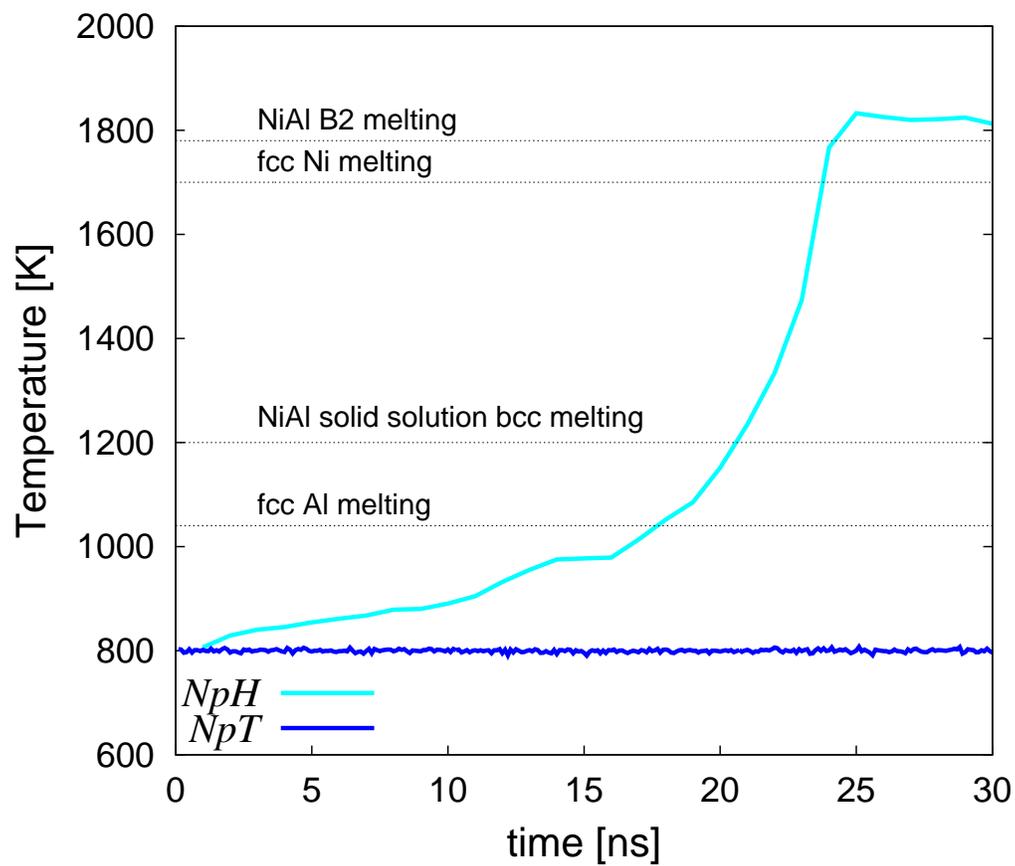}
        \caption{Time evolution of the global temperature for the $NpT$ and $NpH$ simulations. The melting points of the equiatomic NiAl bcc solid solution and ordered B2 phases are shown for reference as obtained from the data in Fig.\ \ref{ff}. The melting temperatures for pure fcc Ni and Al are also shown.}
\label{fig:temp}
\end{figure}

\begin{figure}[h]
        \centering
                \includegraphics[width=\textwidth,trim=1cm 1.0cm 6cm 15.0cm,clip=true]{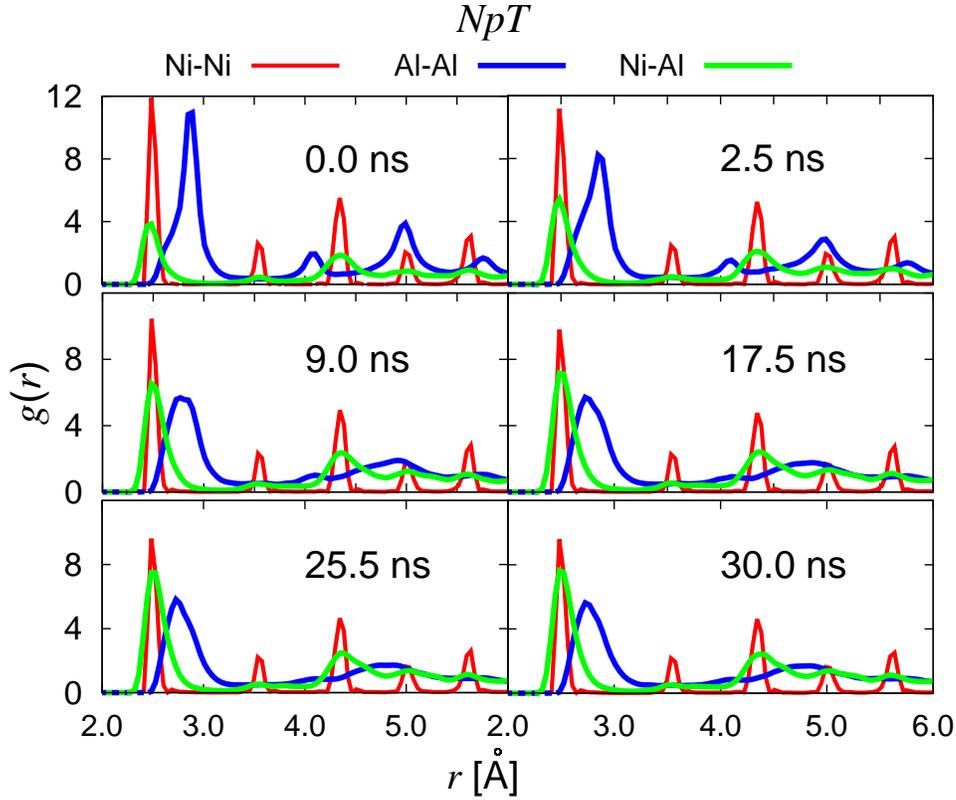}
        \caption{Evolution of $g(r)$ in the $NpT$ ensemble for Ni-Ni, Al-Al, and Ni-Al pairs for each of the snapshots shown in Fig.\ \ref{fig:sequence}. The figures show an evolution from a weakening fcc towards disordered Al-Al correlations and an increasingly stronger bcc Ni-Al signal. The Ni-Ni fcc signal maintains its strength owing to the reservoir of fcc Ni that remains through time.}
\label{fig:gofr}
\end{figure}

A sequence of snapshots from the 30-ns reaction simulation is given in Figure \ref{fig:sequence}. 
Following equilibration, a Nos\'e-Hoover thermo-barostat with a damping constant of 1.0 ps$^{-1}$ is used to maintain a constant temperature of 800 K. The time evolution of the temperature for the entire simulation is shown in Figure \ref{fig:temp}. The initial stages of the reaction process are a continuation of the main features of the sample preparation, namely, swift penetration of Ni atoms in the fcc Al crystal and limited Al diffusion in Ni. Eventually this leads to the formation of an unstructured (amorphous) phase in the Al-rich region, as confirmed by pair correlation function $g(r)$ analysis. $g(r)$ is computed in a 20-\AA~thick slab that encompasses the original interface and its evolution is shown in Figure \ref{fig:gofr}\footnote{Note that, at 800 K, Al is approaching its melting limit and $g(r)$ is quite broad for the fcc structure.}. The results for the crystalline and amorphous phases are consistent with those reported in the literature \cite{zhu2007,izvekov2012}. 

According to the Ni-Al phase diagram \cite{massalski}, the dissolution of Ni in the Al half-crystal at 800 K draws a trajectory in the temperature-concentration space that traverses different Al-rich phases of the alloy system until reaching the equiatomic stoichiometry. These include noncubic structures not captured by the present potential, such as NiAl$_3$ and Ni$_2$Al$_3$. At the same time, there is experimental evidence that melting occurs during reaction of Ni and Al in environments with atomic ratios close to 1:3 \cite{ma1990,zhua2002}. Crystalline NiAl phases subsequently emerge from the melt, giving rise to a stable alloy. As it will be discussed in Section \ref{sec:disc}, simulations in the $NpT$ ensemble prevent melting by thermostatting the exothermic release due to the Ni-Al reaction ($\approx$0.32 eV per atom)\footnote{This result is obtained from Fig.\ \ref{ee} as $\Delta E_r=0.5(E_{\rm{Ni}}+E_{\rm{Al}})-E^{am}_{\rm{NiAl}}$, which, at 800 K, is $\Delta E_r\approx0.5(-4.35-3.19)+4.09=-0.32$ eV per atom.}. 

\begin{figure}[h]
\begin{center}
\includegraphics[width=12cm]{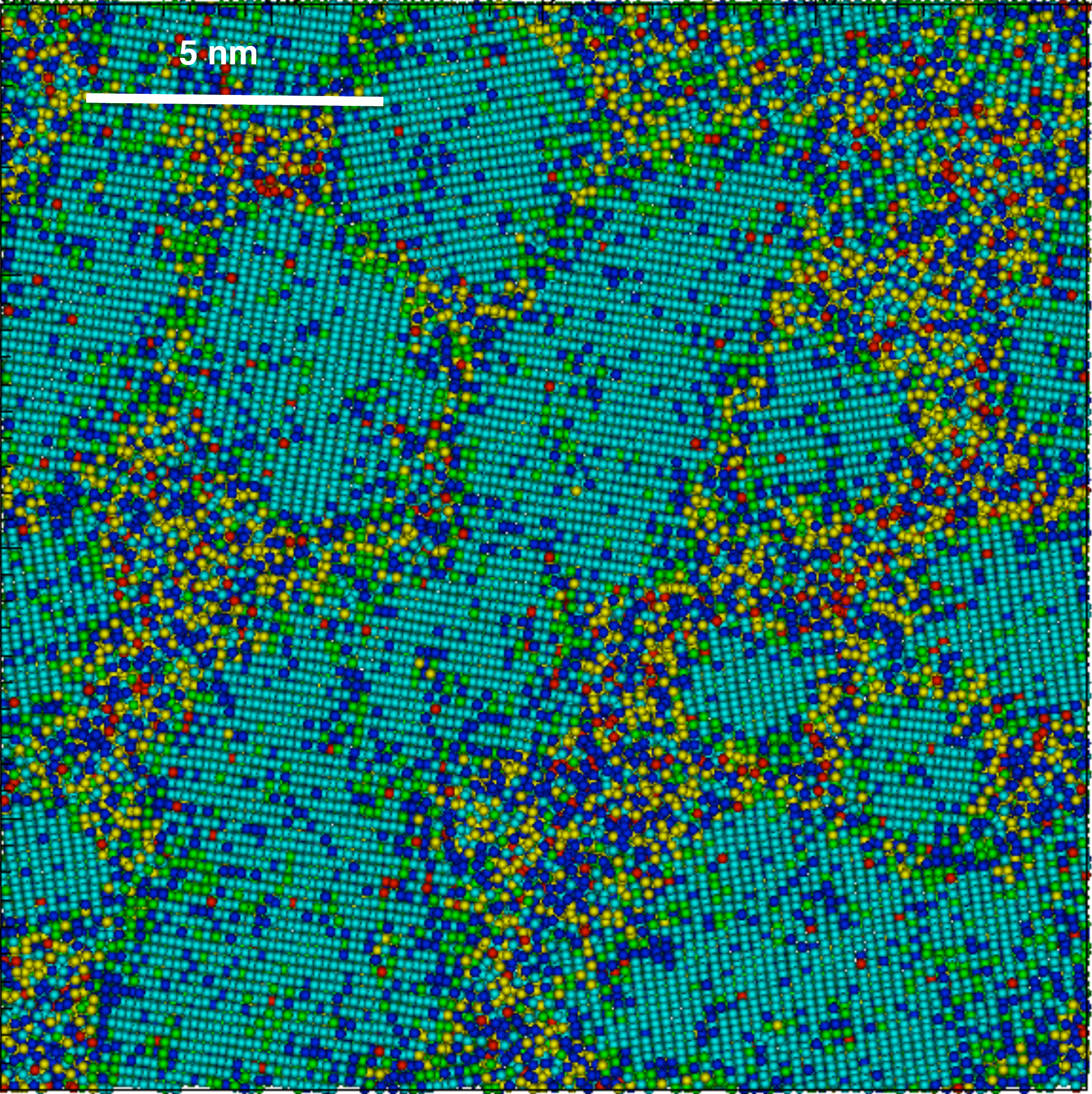}
\caption{Crystal structure at the end of a 30-ns simulation of Ni-Al at 800 K. The image corresponds to a cut parallel to the original Ni-Al interface taken at a distance of 5 \AA~into the Al layer. The observed microstructure consists of several B2 grains of approximately 5 nm in size. The structure was quenched from 800 K to remove thermal noise and make visualization clearer.}
\label{b2_quenched}
\end{center}
\end{figure}

The formation and growth of the unstructured phase continues up to a time of approximately 20 ns in our simulation. From there on,  thermodynamics drives the system toward structures consistent with Fig.\ \ref{ff}, {\it i.e.} B2 phases. Recrystallization of the amorphous phase into a B2 structure initiates at the interface, resulting in a metastable (nano) crystalline structure characterized by high-angle boundaries and uncorrelated grain orientations. Evidence for this nano crystalline B2 structure is provided in Figure \ref{b2_quenched}, which shows a still frame of the final system 30 ns after equilibration taken 5 \AA~from the original interface location on the Al-rich side. 

The image shows atoms colored according to their Ackland-Jones parameter \cite{aj}, which shows the formation of a nano grained bcc structure close to equiatomic composition\footnote{According to the lammps convention, the Ackland-Jones parameter takes values of 0 (blue): unknown, 1 (cyan): bcc, 2 (green): fcc, 3 (yellow): hcp, and 4 (red): icosahedral.}. Therefore, the analysis is conclusive in terms of the final phase formed, and is consistent with the free energy calculations observed in previous sections.

\subsubsection{$NpH$ ensemble}\label{nph}
\begin{figure}[h]
        \centering
        \begin{subfigure}[h]{0.27\textwidth}
                \includegraphics[width=\textwidth]{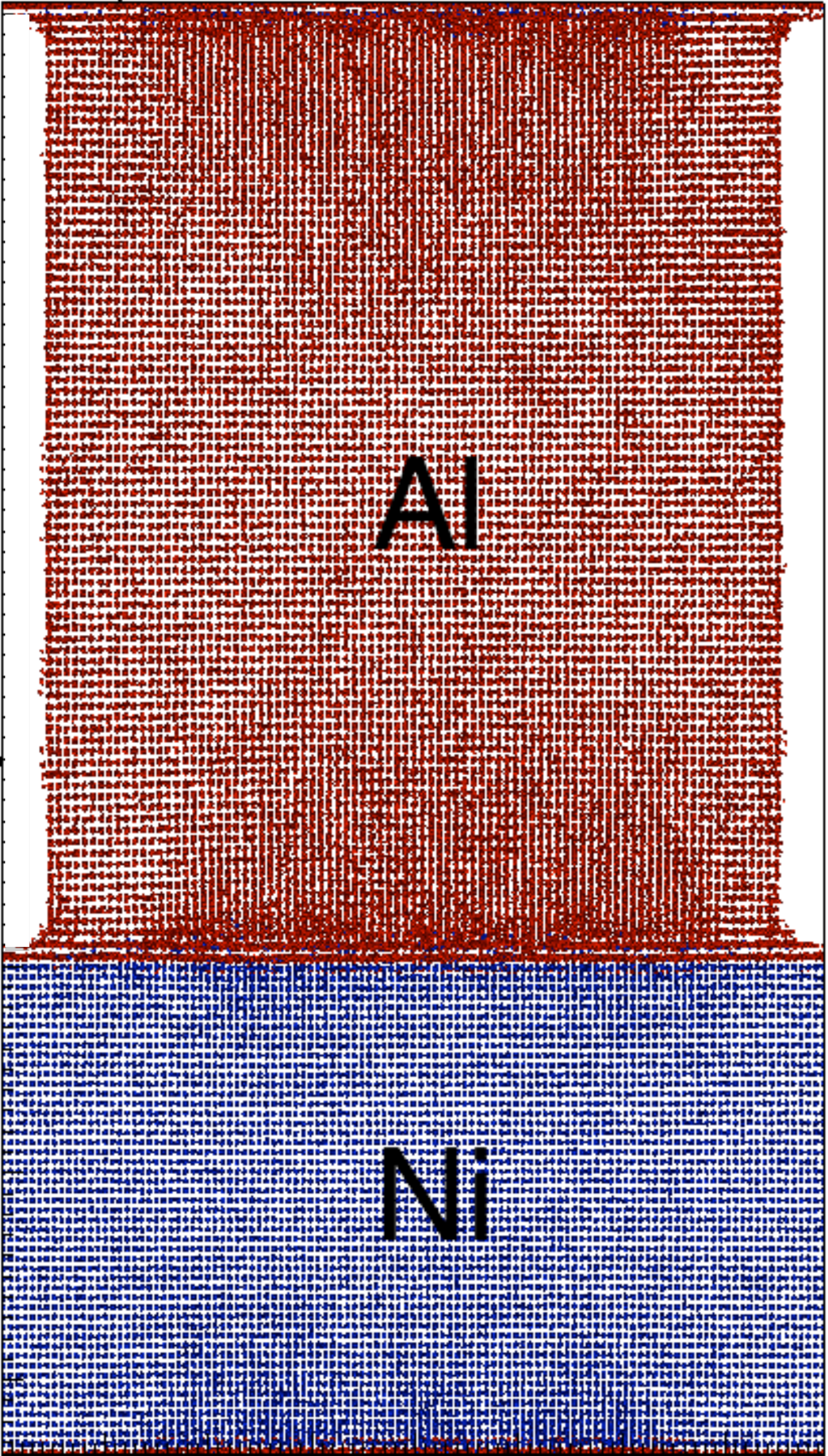}
                \caption{$t=0$ ns}
                \label{framenph1}
        \end{subfigure}~%
        \begin{subfigure}[h]{0.3\textwidth}
                \includegraphics[width=\textwidth]{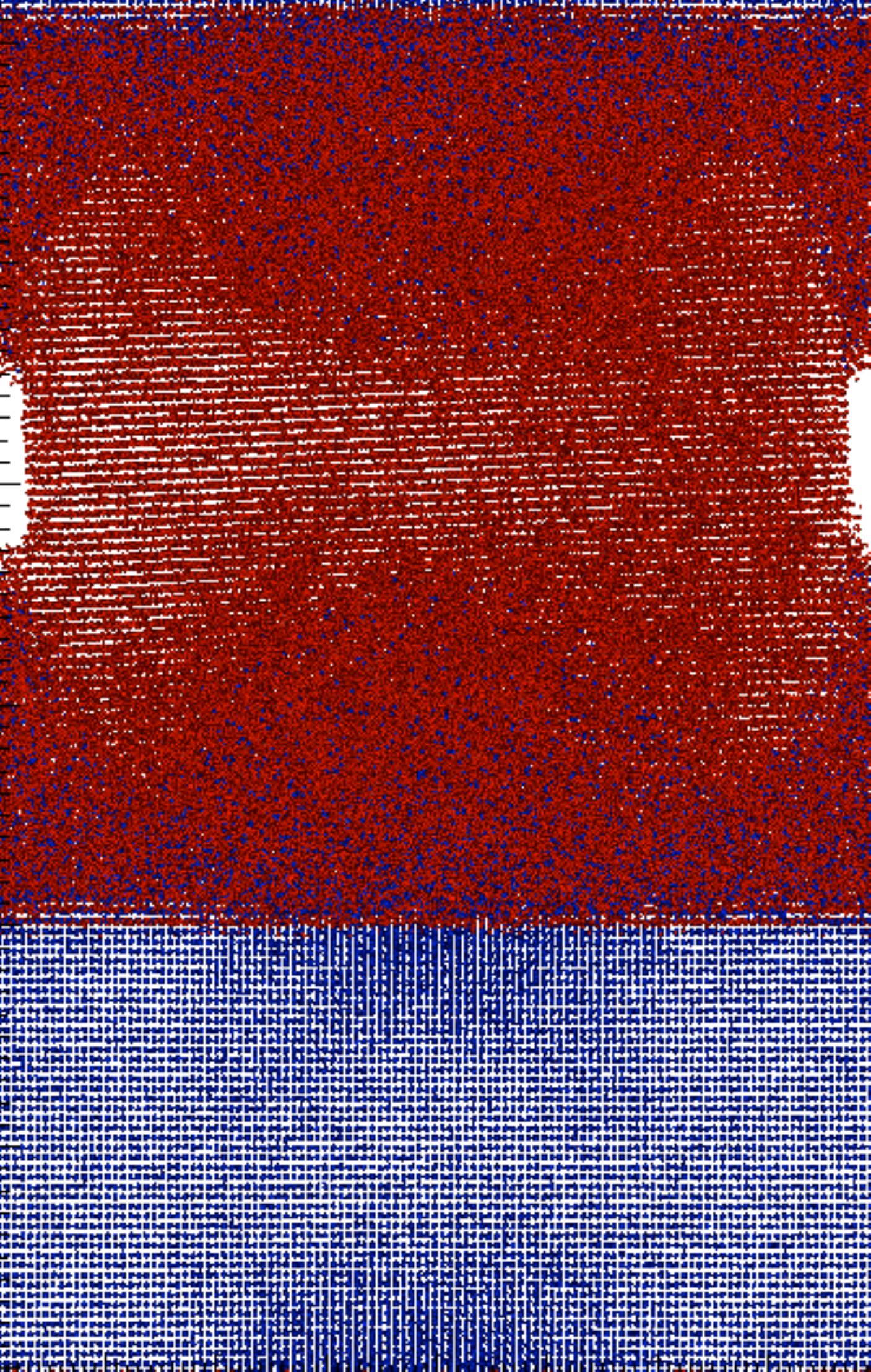}
                \caption{$t=10$ ns}
                \label{framenph2}
        \end{subfigure}
        \begin{subfigure}[h]{0.35\textwidth}
                \includegraphics[width=\textwidth]{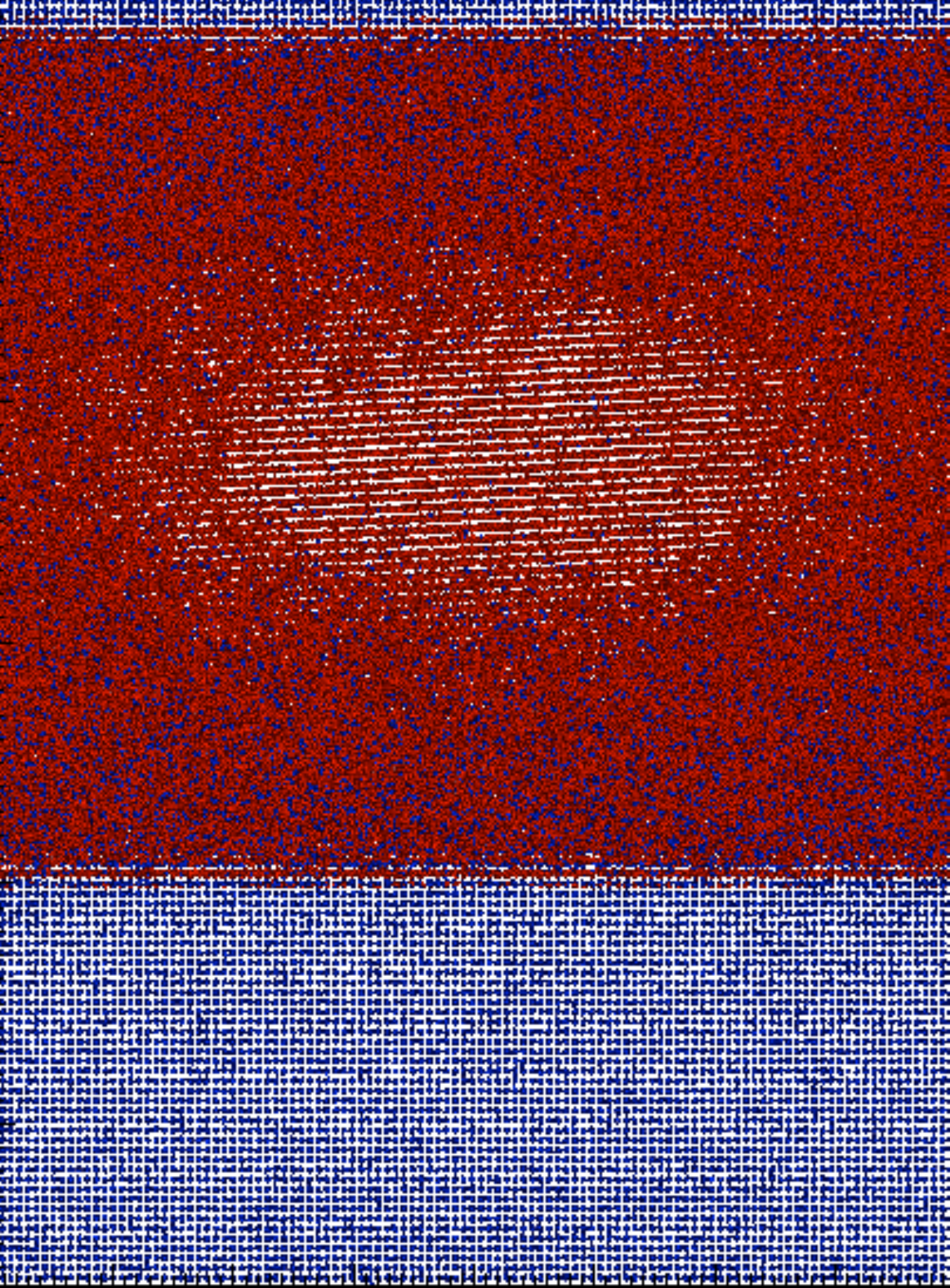}
                \caption{$t=13$ ns}
                \label{framenph3}
        \end{subfigure}
        \begin{subfigure}[h]{0.3\textwidth}
                \includegraphics[width=\textwidth]{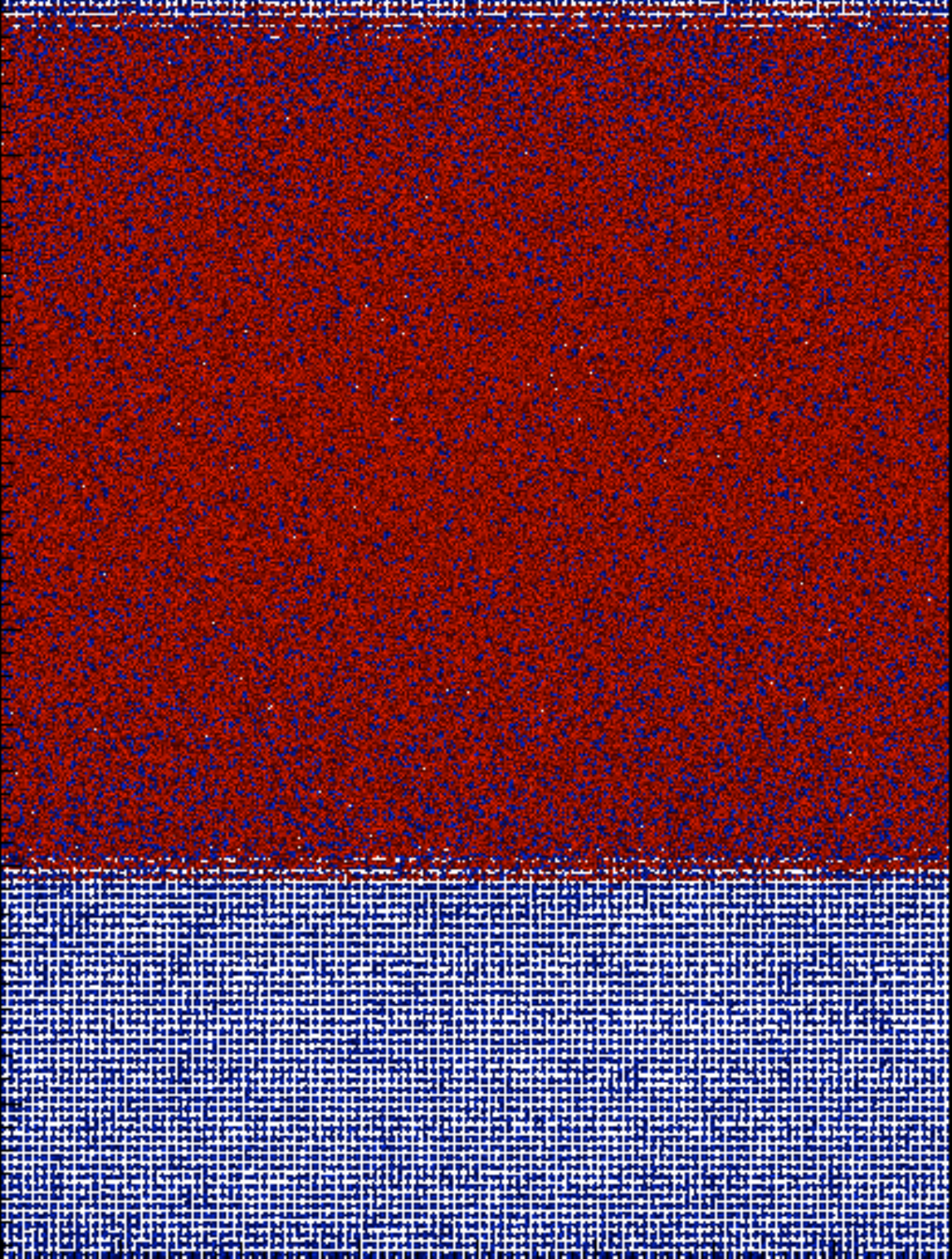}
                \caption{$t=16$ ns}
                \label{framenph4}
        \end{subfigure}
        \begin{subfigure}[h]{0.32\textwidth}
                \includegraphics[width=\textwidth]{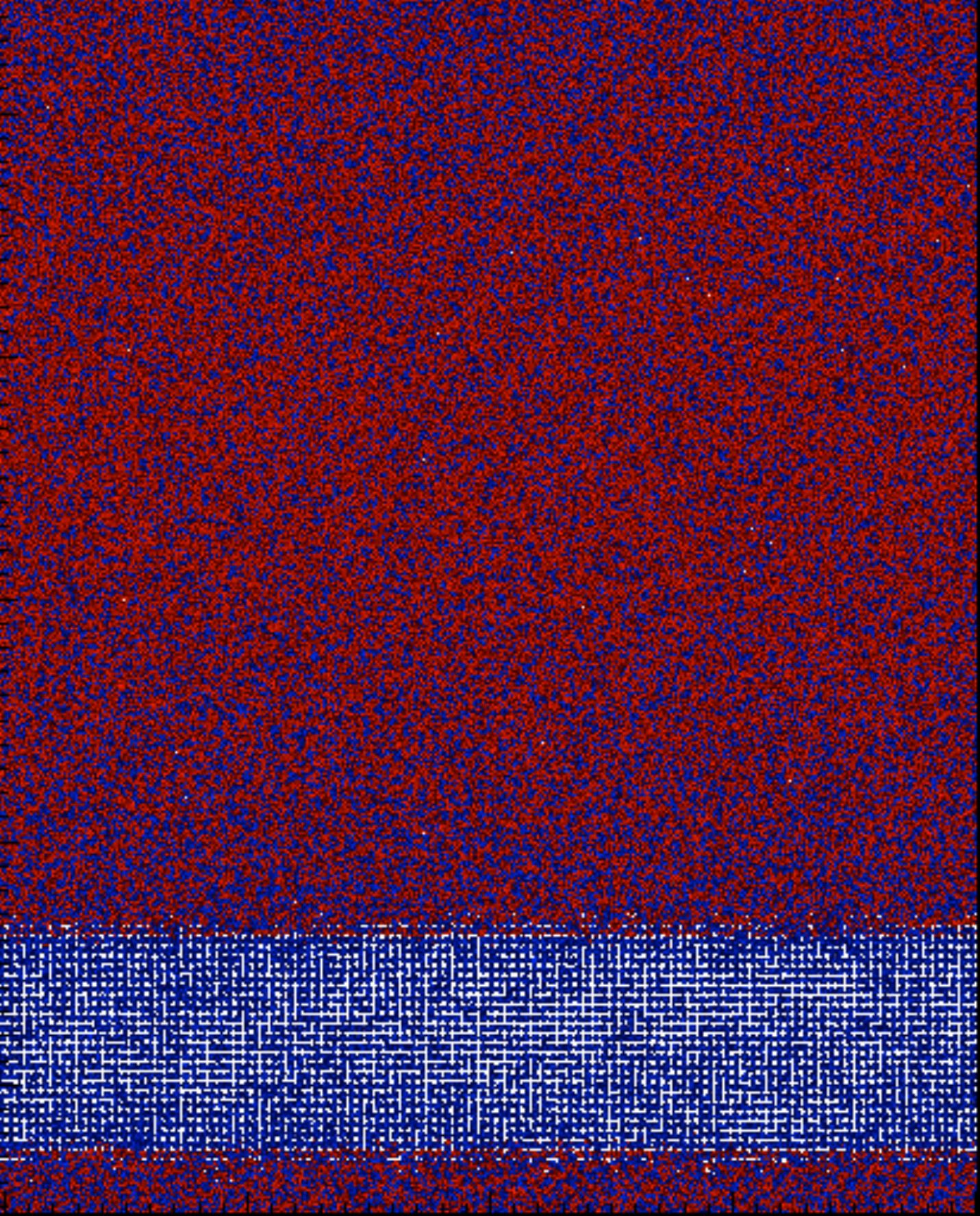}
                \caption{$t=22$ ns}
                \label{framenph5}
        \end{subfigure}
        \begin{subfigure}[h]{0.28\textwidth}
                \includegraphics[width=\textwidth]{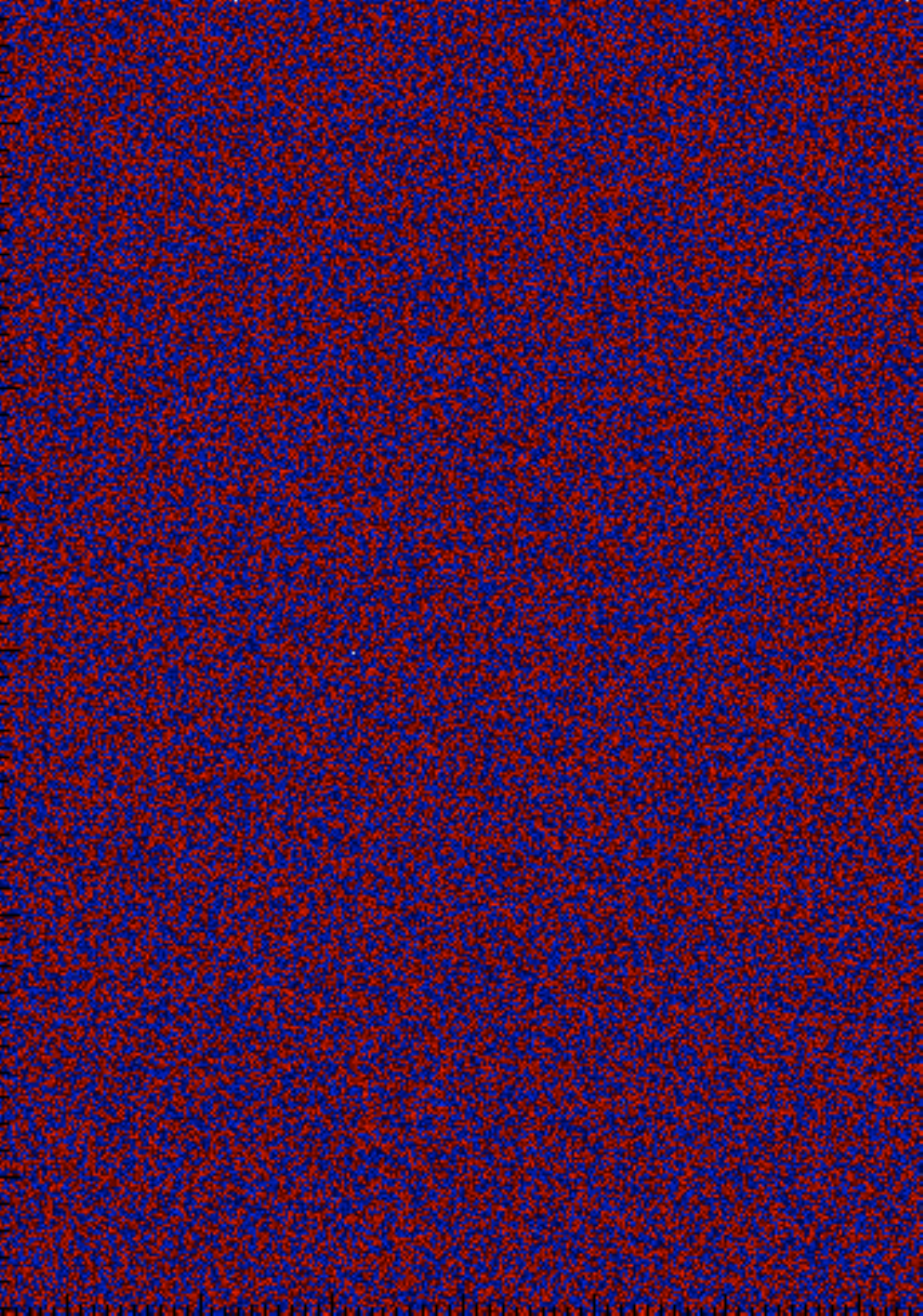}
                \caption{$t=26$ ns}
                \label{framenph6}
        \end{subfigure}
        \caption{Time sequence of the reaction process in a Ni-Al bilayer at 800 K in the $NpH$ ensemble. Red circles represent Al atoms, blue circles symbolize Ni atoms. (a) Initial system during equilibration (identical to Fig.\ \ref{frame1}). (b) The Al half-crystal expands relaxing all stresses. A disordered mixture starts to form. (c) Disorder almost complete. A reaction front propagates into the Ni-rich half-crystal. (d) Full disorder is achieved in the Al half-crystal. (e) Full liquefaction is achieved. Reaction front progresses. (f) Full mixing achieved.}
\label{fig:sequence2}
\end{figure}

\begin{figure}[h]
\begin{center}
\includegraphics[width=\textwidth,trim=1cm 1.0cm 6cm 15.0cm,clip=true]{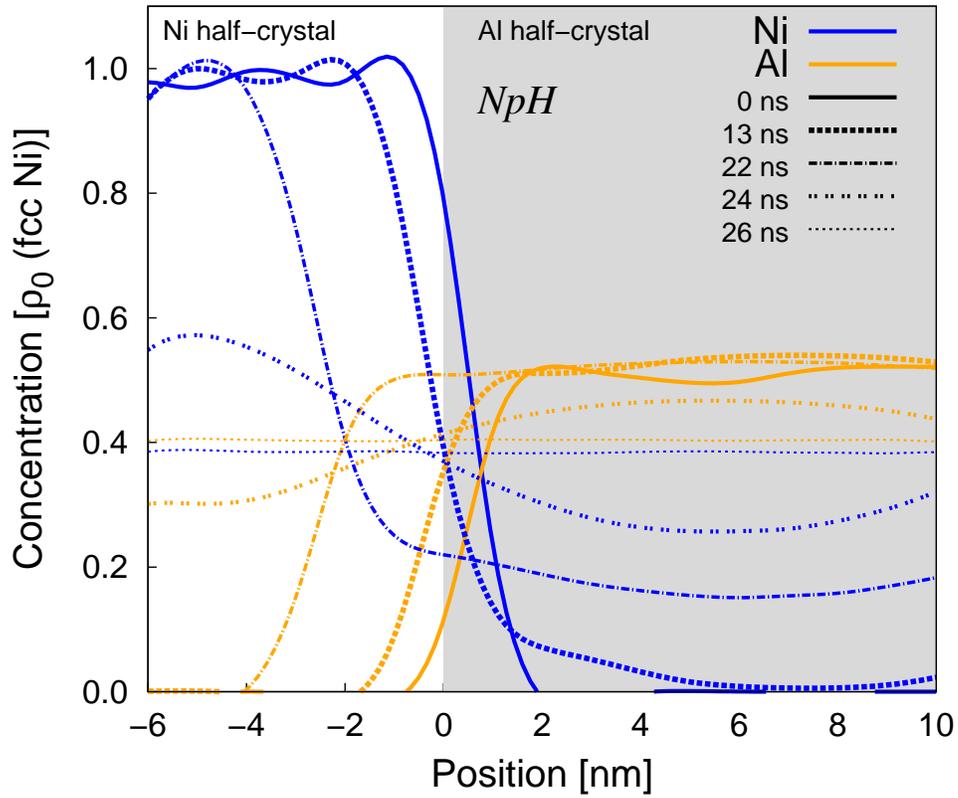}
\caption{Time evolution of the Ni and Al concentration profile corresponding to the bi-crystal reaction simulation in the $NpH$ ensemble. The  concentration is given relative to the fcc Ni atomic density at 0 K (cf.\ Table \ref{table}). The shaded area represents the original extent of the fcc Al half-crystal.}
\label{proff}
\end{center}
\end{figure}

A sequence of snapshots from the 30-ns reaction simulation is given in Figure \ref{fig:sequence2}. The corresponding time evolution of the temperature is given in Figure \ref{fig:temp}.
As for the $NpT$ case, Al diffusion into Ni is practically negligible, whereas Ni atoms quickly penetrate the Al half-crystal. This spurs the formation of a disordered region in the Al side that eventually encompasses the entire half-crystal (Fig.\ \ref{framenph4}, after 16 ns of simulation).  This is confirmed by Figure \ref{proff}, which shows a constant nonzero background Ni concentration in the Al region after 13 ns. This also causes an upturn in the temperature of the box, cf.\ Fig.\ \ref{fig:temp}, which reaches the melting point of the NiAl B2 phase at 24 ns. This is followed by a propagation of the interface between the pure fcc Ni crystal and the Ni-Al disordered region, resulting in a gradual increase of the relative Al concentration in the original Ni region.  By 26 ns the entire computational box has transformed into a liquid with the temperature remaining constant at approximately 1850 K. 

\begin{figure}[h]
\begin{center}
\includegraphics[width=\textwidth,trim=1cm 1.0cm 6cm 15.0cm,clip=true]{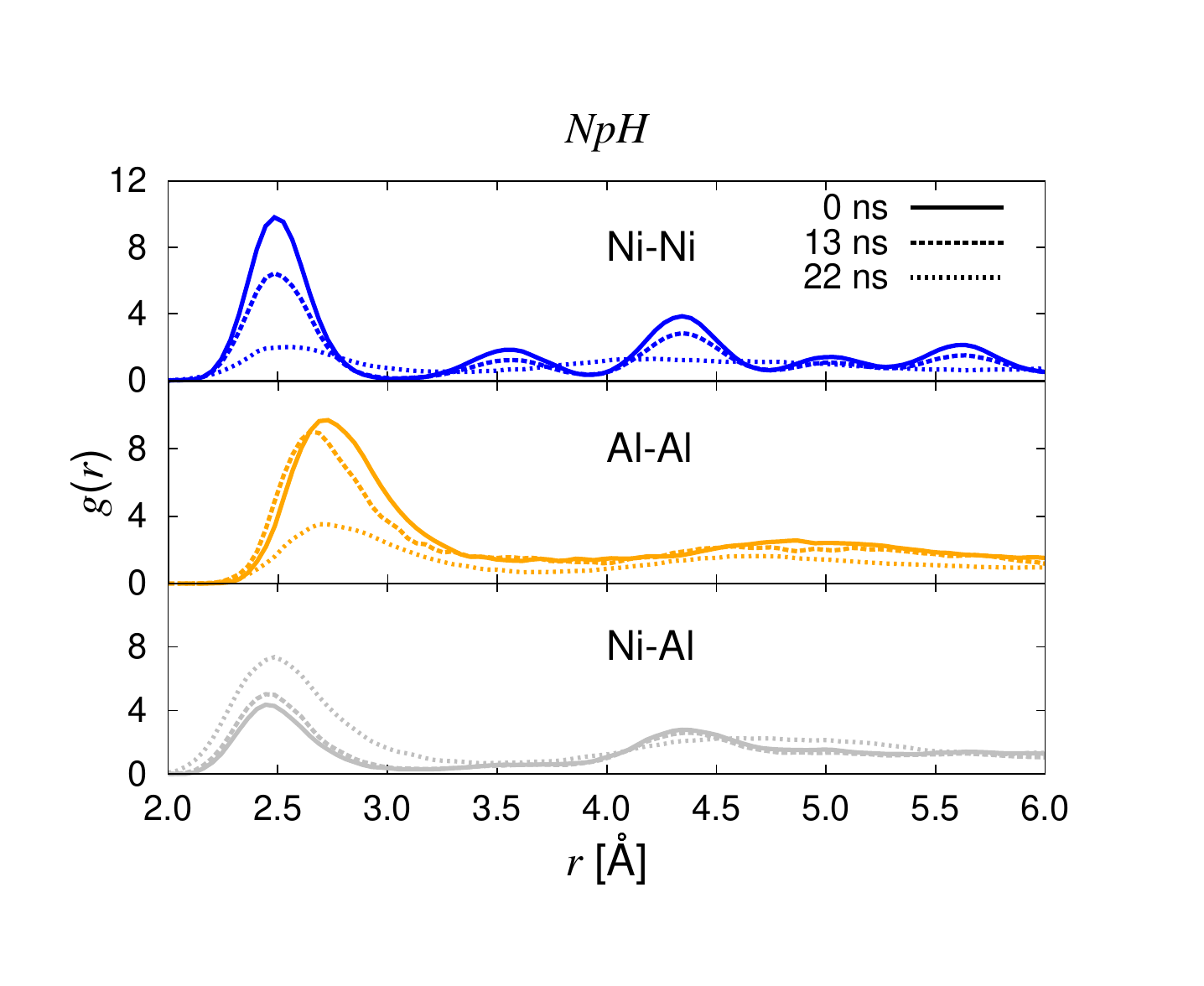}
\caption{Pair correlation function as a function of time for the reaction process simulated in the $NpH$ ensemble.}
\label{gofrnph}
\end{center}
\end{figure}

The above description is substantiated by the analysis of the time evolution of the $g(r)$ function. This is shown in Figure \ref{gofrnph}, which confirms the existence of decreasing tendencies for Ni and Al fcc pairs and an increasing one for disordered Ni-Al pairs.

\section{Discussion}\label{sec:disc}
\subsection{Thermodynamics of Ni-Al compounds} 
Results in Table \ref{table} match, where appropriate, those obtained by Purja and Mishin \cite{mishin2009}. Our calculations do not include zero-point motion at low temperatures and are thus technically only suited for temperatures above the Debye temperature ($\approx$430 and 450 K for Al and Ni, respectively). However, our data are in reasonable agreement with those reported by Wang \etal~\cite{wang2004} at low temperatures using first principles calculations. Asta \etal~\cite{asta1999} calculated several structural and physical properties of liquid Ni-Al mixtures using different interatomic potentials, and concluded that in, mixtures rich in Al ($c<0.5$), interatomic potentials of the embedded-atom type as the one employed here underestimate the viscosity and diffusivity relative to available experimental data. However, we do not consider this example to be fully representative of our calculations, as it pertains to liquid phases with different concentrations. 

Besides these studies, there are to our knowledge no studies concerning the thermodynamic properties of Ni-Al systems in such an extensive temperature range.

\subsection{Ni-Al Reaction kinetics and thermodynamics}
Ni-Al mixing processes have been studied in the literature using atomistic simulations with a combination of $NVE$ and $NpH$ ensembles\cite{henz2009,weingarten2010,ale2011,ale2012,izvekov2012}.
In our case, it is important to discuss our simulations in the context of the exothermic heat release due to the Ni-Al reaction. As mentioned in Section \ref{sebsec:kin}, the excess heat resulting from the formation of amorphous (liquid) NiAl relative to isolated Ni and Al at 800 K is approximately 0.32 eV per atom. In the absence of any mechanical work performed on the system, this would result in a temperature increase of:
$$\Delta T\approx\Delta E_r/C_p$$
which, taking the values for the heat capacity from Table \ref{table}, would results in temperature increases on the order of 1230 K. This would suggest a final temperature of $\approx$2030 K\footnote{Of course, once the melting point of the most stable thermodynamic phase is reached ---the B2 phase at 1780 K---, the released heat is invested as latent heat and the temperature does not change.} at the end of the simulations and, as discussed in Section \ref{sebsec:kin}, melting of all the phases involved.
This picture is in agreement with other MD simulations using the same interatomic potential \cite{ale2012} in the $NpH$ ensemble. 

This is evidently not what happens during the $NpT$ simulations, which resemble an infinitely slow process where the released heat has sufficient time to dissipate and the temperature is kept constant. In such a case, one would expect the system to display the equilibrium phase at each composition. 
The reaction process in the $NpT$ ensemble can be understood as following a rectilinear path along the 800-K isotherm on the $F$-$T$ plane shown in Fig.\ \ref{ff} (this path is shown descriptively in the figure). The reaction process occurs via the formation of a mixing zone of diffuse nature, characterized by the penetration of Ni in the Al layer, that grows as Ni diffuses and reaches concentrations capable of resulting in stable Ni-Al compounds. There is atomistic evidence that the minimum temperature for the process to occur in this fashion is of the order of 700 K \cite{zhang2011}. 
Because Ni interpenetration in Al is much larger than vice versa, the bilayer reaction can be thought of as a process where Ni arrives in the Al crystal and gradually increases its relative concentration from zero to 0.5. Thermodynamically, this is illustrated also in Fig.\ \ref{fig:comp} via the chemical potential difference $\Delta\mu=\partial F/\partial c$ (where $c$ is the Ni concentration), which is negative for all values of $T$ up to $c\approx0.8$. This trajectory means that any incremental addition of Ni results in a decrease of $F$ and is therefore thermodynamically favored. However, in going from low Ni concentrations to $c\approx0.5$ along the 800-K isotherm (or, in fact, any other), one crosses a region of the free energy surface where the amorphous phase is more stable than the bcc phase (at 800 K, between $c\gtrsim0.08$ and $c\lesssim0.40$). This is essentially what happens between 9 and 25 ns after the reaction initiation, and signals the local melting of the bcc solid solution at that composition and temperature. As the Ni concentration continues to increase, the system recrystallizes again until an equiatomic compound is formed. This occurs first near the Ni-Al interfaces, which spur the nucleation of a lowest-free-energy B2 phase. Interestingly, a nano grained structure appears as a result of the impingement of several independently nuclei. The constitution of the bcc and B2 phases takes place according to classical (heterogeneous) nucleation, and thus, although this case may not necessarily be considered as \emph{realistic}, it is thermodynamically consistent in terms of the free energy calculations reported here.  

For their part, $NpH$ simulations reveal a very different picture. In this case, the temperature does increase up to the melting point of the B2 phase, creating a liquid phase that encompasses the entire simulation box. Thermodynamically, the simulations may be regarded as following a constant $H\equiv E$ path in Fig.\ \ref{ee}: at 800 K, $(E_{\mathrm{Ni}}+E_{\mathrm{Al}})/2\approx-3.72$ eV/atom. Following this isenthalpy along the temperature axis, one can see that the liquid (amorphous) phase is encountered at $\approx$1800 K, in good agreement with the melting point of the B2 structure. The path along temperature is not uniform but, as Fig.\ \ref{fig:temp} shows, one that is gradual up to the melting point of Al and that steepens thereafter until an absolute liquid is formed. The rate of temperature increase during the first stage is approximately 15 K ns$^{-1}$, while for the second one it is 60 K ns$^{-1}$. This accelerated exothermic release correlates with a homogeneous mixture of Ni-Al in the original Al half crystal. It is at this point that the interface begins to move into the original Ni half crystal, resulting in rapid heat production. The full melting point is achieved when the interface traverses the remaining crystalline zone. The $NpH$ simulations result in a net temperature increase of $\Delta T\approx1050$ K. Following the above reasoning, this is the result of $C_p\Delta T\approx0.27$ eV per atom released into the system. Because the total released energy is 0.32 eV per atom, this implies that $0.32-0.27=0.07$ eV per atom are released as latent heat.

\begin{figure}[h]
\begin{center}
\includegraphics[width=\textwidth]{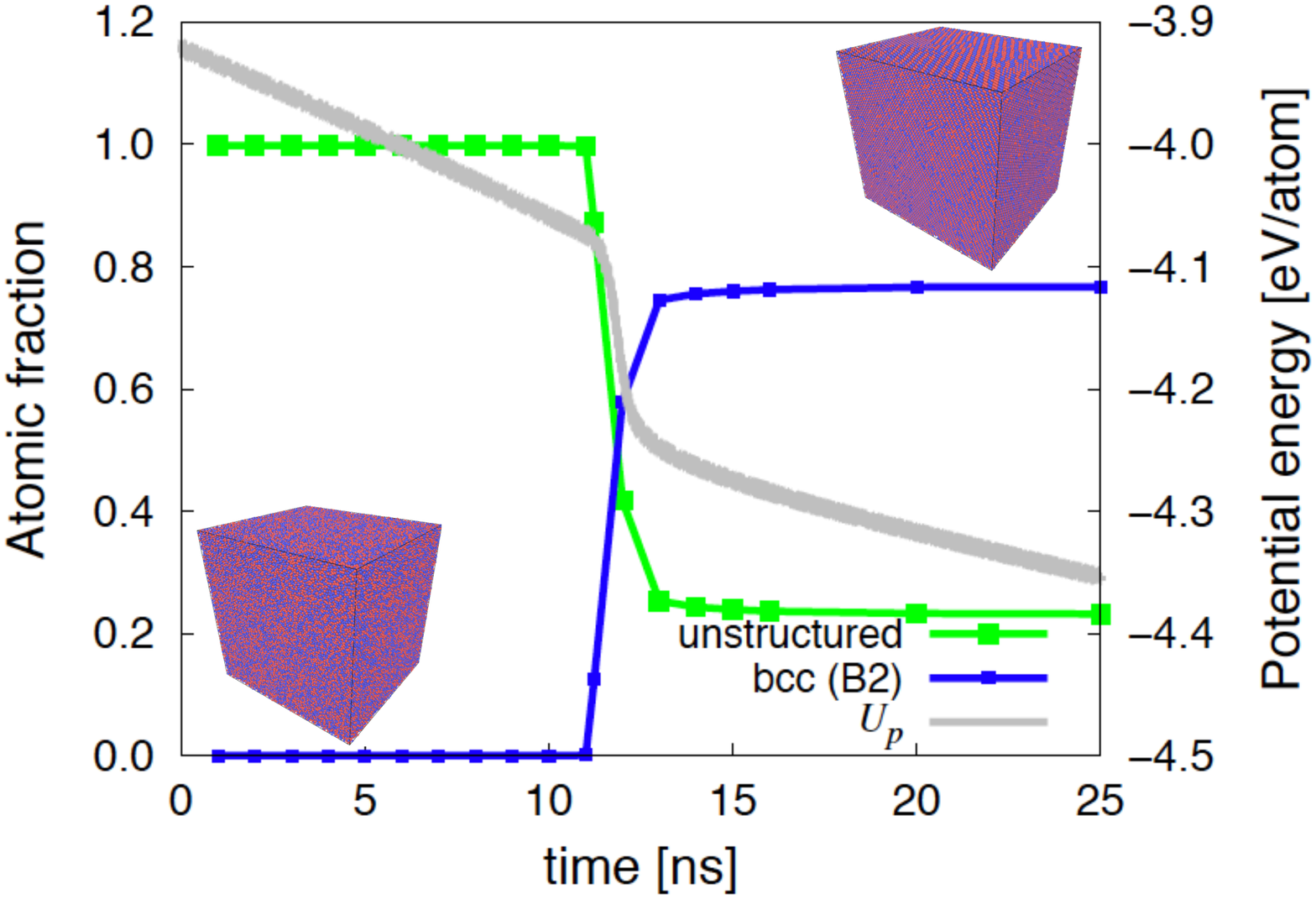}
\caption{Time evolution of the fraction of atoms with a particular atomic structure for the cooling of the liquid $NpH$ configurations (Fig.\ \ref{framenph6}). The concomitant evolution of the system's potential energy is also shown. The phase transformation occurs at 11 ns of simulation time, which corresponds to a temperature of 1250 K. The figure includes two snapshots of the configurations at times 1 and 20 ns (Al: red atoms; Ni: blue atoms), where the temperatures are, respectively, 1800 and 800 K.}
\label{cool}
\end{center}
\end{figure}

In experimental conditions, the exothermic heat released during this reaction would dissipate over time, cooling the liquid configuration below the melting point and resulting in a stable crystalline phase. To study whether this phase, as in the $NpT$ case, is B2, we cool the liquid configuration corresponding to the end point of the 30-ns $NpH$ MD simulation down to 600 K at a cooling rate of 50 K ps$^{-1}$. Figure \ref{cool} shows the evolution with time of the fraction of atoms with different crystalline structure during the cooling $NpH$ simulation. As shown, approximately after 11 ns, corresponding to a temperature of 1250 K, the system phase transforms into a B2 structure. The figure shows a discontinuous transition of the system's potential energy, which suggests a first-order phase transformation. For the B2 phase, this crystallization temperature of 1250 K must be compared with the melting point of 1780 K to give an idea of the degree of hysteresis achieved in the simulations.
The fraction of `unstructured' atoms before this transformation represents the liquid structure while, after it, it corresponds to grain boundaries and local disordered zones interspersed in the B2 structure.

It must be noted that, to first order, the time scale for heat dissipation can be obtained by assuming that the system dissipates heat at a rate given by Newton's law of cooling:
$$T(t) = T_0\exp\left(-\frac{t}{\tau}\right)$$
where $T_0$ is the temperature of the heat bath, and:
$$\tau=\frac{mC_p}{\kappa L}$$
is the time constant associated with this process, where $\kappa$ is the thermal conductivity and $L$ is a length scale that represents the distance from the heat source to the bath. Taking values from Table \ref{table} for the heat capacity and from the literature for $\kappa$ \cite{kappa}, we obtain a value of $\tau=87$ ps, well within the times simulated with MD here. This implies that, after a few hundred ps, the system would start to dissipate heat to a heat bath if there were one available. However, on the one hand, this value of $\tau$ is not sufficiently low to be used as damping constant in the $NpT$ simulations (cf.\ Ref.\ \cite{lammps}). On the other, it means that heat would start to flow out of the system within the time scale of the $NpH$ simulations, which is not a process captured here. In any case, we provide this discussion to give the reader an idea of the realism of the simulations done here.


\section{Conclusions}
We have obtained the thermodynamic properties of intensive thermodynamic variables for the N-Al system as a function of structure and composition.
We have performed atomistic simulations of the Ni-Al system at high homologous temperatures and extracted several thermodynamic quantities for these conditions. This simulation methodology was then applied to a Ni-Al diffusion couple and its evolution observed. The simulations were done at 800 K and zero pressure under $NpT$ and $NpH$ conditions. In both cases, the Ni atoms diffuse quickly into the Al and this alloying causes a structural transformation to an amorphous phase. This amorphous region corresponds to a highly disordered Ni-Al mixture in the $NpT$ ensemble, while it represents the absolute liquid in $NpH$ simulations. In the former case, amorphous regions near the interface ---where the composition first reaches the 1:1 ratio--- nucleate into grains of the B2 Ni-Al intermetallic phase. These grow and collide with one another, resulting in a nano crystalline B2 structure. In the latter case, a gradual cooling of the liquid phase also results in a B2 system. 
The absence of other intermetallic phases such as NiAl$_3$ or Ni$_3$Al may at first be counterintuitive because these compositions will at some point be present upon cooling of a liquid or recrystallization of an amorphous phase. However, the free energy calculations indicate that the B2 phase has a higher thermodynamic driving force for formation that leads to more rapid kinetics. These other intermetallic phases are also not observed experimentally in systems with the small length scales and rapid kinetics considered here \cite{kim2008}.

\section*{Acknowledgments}
We thank Dr.~A. Caro for critically reviewing the manuscript. This work was performed under the auspices of the US Department of Energy by Lawrence Livermore National Laboratory under Contract DE-AC52-07NA27344. The contributions of LS and GHC to this work were supported by DOE Office of Science, Office of Basic Energy Sciences, Division of Materials Sciences and Engineering. JM acknowledges support from the DOE's Early Career Research Program.

\end{document}